\documentclass[12pt]{article}
\usepackage{epsfig}
\usepackage{amsfonts}
\usepackage{latexsym}
\usepackage{amsmath}
\usepackage{mathrsfs}
\usepackage{hyperref}
\usepackage{wasysym}
\usepackage{tikz}
\usepackage{caption}
\usepackage{subcaption}
\usepackage{float}
\numberwithin{equation}{section}

\DeclareFontFamily{OT1}{rsfs}{}
\DeclareFontShape{OT1}{rsfs}{m}{n}{
<-7> rsfs5 <7-10> rsfs7 <10-> rsfs10}{}
\DeclareMathAlphabet{\mycal}{OT1}{rsfs}{m}{n}

\newcommand{\be}{\begin{equation}}
\newcommand{\ee}{\end{equation}}
\newcommand{\bea}{\begin{eqnarray}}
\newcommand{\eea}{\end{eqnarray}}
\def\non{\nonumber}
\def\pa{\partial}
\def\d{\delta}

\def\e{\text{\dh}}
\def\t{\text{\thorn}}
\def\p{\psi}
\def\D{\Delta}
\def\non{\nonumber}

\def\a{\alpha}
\def\tc{\tilde{\chi}}

\title{Is there a breakdown of effective field theory at the horizon of an extremal black hole?}

\author{Shahar Hadar and Harvey S. Reall \\{\small Department of Applied Mathematics and Theoretical Physics,} \\ {\small
 University of Cambridge, Wilberforce Road, Cambridge CB3 0WA, UK}}

\begin{document}

\maketitle

\unitlength = 1mm

\setcounter{tocdepth}{2}

\begin{abstract}
Linear perturbations of extremal black holes exhibit the Aretakis instability, in which higher derivatives of a scalar field grow polynomially with time along the event horizon. This suggests that higher derivative corrections to the classical equations of motion may become large, indicating a breakdown of effective field theory at late time on the event horizon. We investigate whether or not this happens. For extremal Reissner-Nordstrom we argue that, for a large class of theories, general covariance ensures that the higher derivative corrections to the equations of motion appear only in combinations that remain small compared to two derivative terms so effective field theory remains valid. For extremal Kerr, the situation is more complicated since backreaction of the scalar field is not understood even in the two derivative theory. Nevertheless we argue that the effects of the higher derivative terms will be small compared to the two derivative terms as long as the spacetime remains close to extremal Kerr.
\end{abstract}

\pagestyle{plain}
\setcounter{page}{1}
\newcounter{bean}
\baselineskip18pt


\setcounter{tocdepth}{2}


\section{Introduction} \label{Introduction}

Extremal black holes (BHs) are an important special class of BHs with degenerate, zero temperature horizons. They play a prominent role in String Theory as they are often supersymmetric and do not evaporate. As distinguished members of the BH family with broad theoretical applications, understanding their classical stability properties seems important. Are extremal BHs classically stable?

While proving the nonlinear stability of the Kerr BH remains as a major goal of mathematical relativity, some significant steps towards this goal have already been made. The current state-of-the-art are the recent proofs of linear stability of Schwarzschild under gravitational perturbations \cite{Dafermos:2016uzj} and linear stability of a massless scalar on Kerr \cite{Dafermos:2014cua}. Importantly, these proofs are restricted to non-extremal BHs. The reason is that the so-called horizon redshift effect is essential in those analyses. This is the phenomenon that outgoing radiation propagating along the future event horizon suffers a redshift and therefore decays. The characteristic decay time is proportional to the BH's surface gravity. At extremality the surface gravity vanishes so there is no horizon redshift effect and the stability proofs fail.

The search for a new approach to study the stability of extremal BHs led Aretakis, in a series of works \cite{Aretakis:2011ha,Aretakis:2011hc,Aretakis:2012ei,Aretakis:2012bm}, to prove that massless scalar perturbations of extreme Reissner-Nordstr{\"o}m (RN) and axisymmetric massless scalar perturbations of extreme Kerr BHs display both stable and unstable properties. He showed that the scalar field and its derivatives decay outside the event horizon. However, on the event horizon, the absence of a horizon redshift effect means that outgoing radiation propagating along the event horizon does not decay. Mathematically, this means that a transverse derivative of the scalar field does not decay along the horizon and higher transverse derivatives grow with time. For spherically symmetric massless scalar perturbations of extreme RN, derivatives blow up at least as fast as
\bea
\left. \pa^{k}_r \p \right|_{\mathrm{horizon}} \sim v^{k-1} \, \, ,
\label{derivative behaviour on horizon intro}
\eea
where $\p$ is the field under study, and $(v,r)$ are ingoing Eddington-Finkelstein coordinates. An important element of Aretakis' work is the identification of an infinite set of conserved quantities, along the event horizon, one for each spherical harmonic. These are called the Aretakis constants.

Aretakis' result has been generalized in various ways. Ref. \cite{Lucietti:2012sf} explained why this massless scalar instability afflicts {\it any} extreme black hole, and showed that there is a similar instability for linearized gravitational perturbations of extreme Kerr. Ref. \cite{Lucietti:2012xr} showed that there is a similar instability for coupled gravitational and electromagnetic perturbations of extreme RN, and also for massive scalar perturbations of extreme RN.

The blowup \eqref{derivative behaviour on horizon intro} is `mild' in the sense that it is polynomial rather than exponential. In a frequency domain analysis it therefore appears as a branch point located precisely on the real-frequency axis, rather than as a pole. This was studied recently for extremal Kerr \cite{Casals:2016mel}, and its near-extreme counterpart \cite{Gralla:2016sxp}. It should be noted that these frequency domain analyses cannot describe situations in which there is outgoing radiation initially present at the event horizon. This implies that the results are restricted to cases with vanishing Aretakis constants. With vanishing Aretakis constants there is still an instability but it requires one more derivative to see it \cite{Aretakis:2012bm}, which is precisely what was found in Ref. \cite{Casals:2016mel}.

Ref. \cite{Casals:2016mel} also considered {\it non}-axisymmetric massless scalar perturbations of extreme Kerr and found that they exhibit even worse behaviour than the axisymmetric perturbations considered by Aretakis. Specifically, it was argued that, for non-axisymmetric perturbations, the first transverse derivative of the scalar can grow as $v^{1/2}$ along the horizon (where $v$ is a Killing time coordinate). In \cite{Zimmerman:2016qtn} an extension to charged perturbations of extreme RN was discussed; these were shown to resemble non-axisymmetric modes in extreme Kerr.

The above discussion concerns {\it linear} perturbations of extreme BHs. It is natural to ask what happens when one considers nonlinearity and backreaction. Aretakis considered the case of a scalar field with a particular kind of self-interaction and found that the nonlinearity made the instability worse, leading to a blow up in finite time along the event horizon \cite{Aretakis:2013dpa}. A different kind of nonlinearity was considered in Ref. \cite{Angelopoulos:2016hln}, for which it was found that the nonlinearity did not lead to any qualitative difference from the linear equation. However, for both of these examples, the nonlinearity was not of a kind that would arise in physical applications. The backreaction problem was investigated numerically in Ref. \cite{Murata:2013daa}. It was found that, for a generic (massless scalar field) perturbation, an extreme RN black hole will eventually settle down to a non-extreme RN solution. However, during the evolution, there is a long period when derivatives exhibit the behaviour (\ref{derivative behaviour on horizon intro}), confirming that the instability persists when backreaction is included. Furthermore, by fine-tuning the perturbation it can be arranged that the late-time metric approaches extreme RN, in which case the nonlinear solution exhibits the behaviour (\ref{derivative behaviour on horizon intro}) indefinitely.

We now turn to the physical relevance of the Aretakis instability. If fields decay outside the event horizon then why does it matter that higher transverse derivatives blow up on the horizon? One reason is that we expect the classical equations of motion to be corrected by higher derivative terms, as is the case in string theory. If higher derivatives become large on the horizon then it seems likely that the higher derivative terms in the equation of motion will become large \cite{Murata:2013daa}. In other words, the Aretakis behaviour suggests a possible breakdown of effective field theory at late time on the event horizon of an (arbitrarily large) extreme black hole.\footnote{
A possible late-time breakdown of effective field theory at an event horizon, due to a "string spreading" effect, has been investigated in Ref. \cite{Dodelson:2015toa}. Since this effect is present for non-extremal black holes, it does not appear to be related to the effects discussed in the present paper.}

The aim of this paper is to investigate whether or not higher derivative corrections to the equations of motion become important during the Aretakis instability or the even worse non-axisymmetric extremal Kerr instability of Ref. \cite{Casals:2016mel}. We will consider a nonlinear theory consisting of Einstein-Maxwell theory coupled to a massless scalar, and then add higher derivative corrections which are restricted only by the requirement of general covariance and a shift symmetry for the scalar field.

In section \ref{ERN} we consider the extremal RN solution. We start with a brief review of the Aretakis instability. We then consider the $AdS_2 \times S^2$ near horizon geometry of an extremal RN black hole, taking into account the higher derivative corrections to the background geometry. We expand on a previous discussion \cite{Lucietti:2012xr} of how the Aretakis instability can be seen in the near-horizon geometry. We then show that, for a large black hole, {\it linear} higher derivative corrections lead only to small corrections to Aretakis' results. In particular, the leading (spherically symmetric) instability of the near-horizon geometry is unaffected by these corrections. Ultimately the reason for this is that the higher derivative terms must exhibit general covariance, which implies that they take a very simple form when linearized around a highly symmetric background such as $AdS_2 \times S^2$.

It is not obvious that this will remain true when we consider the much less symmetric geometry of the full black hole solution. So next we consider the size of (possibly nonlinear) higher derivative terms in all of the equations of motion during the Aretakis instability in the full extreme RN geometry. We argue that such terms remain small compared to the nonlinear 2-derivative terms. Hence there is no indication of any breakdown of effective field theory for extreme RN.  Ultimately this result can again be traced back to general covariance restricting the possible form of the higher derivative terms.

In section \ref{EK} we discuss the case of extremal Kerr. Again we start by investigating the scalar field instability in the near-horizon geometry. In particular, we give a simple derivation of results analogous to those of Ref. \cite{Casals:2016mel} for the scalar field instability in the near-horizon extreme Kerr (NHEK) geometry. We explain how these results are robust against higher derivative corrections of the NHEK geometry. Furthermore, our method can incorporate outgoing radiation at the event horizon in the initial data, unlike the approach of Ref. \cite{Casals:2016mel}. Nevertheless, our results are in agreement with those of Ref. \cite{Casals:2016mel}, indicating that this initial outgoing radiation does not make the dominant (non-axisymmetric) instability any worse. We then consider linear higher derivative corrections to the equation of motion for the scalar field and argue that these just give small corrections to the results, again without making the instability any worse. So, at the level of the near-horizon geometry, there is no sign of any breakdown of effective field theory.

Finally we consider the scalar field instability in the full extreme Kerr geometry. Here the effect of nonlinearities is not yet understood, even in the 2-derivative theory. So we simply assume, in analogy with the nonlinear extreme RN results, that the geometry remains close to extreme Kerr even when 2-derivative nonlinearities are included. With this assumption we estimate the size of higher derivative corrections to the equations of motion. We find that these remain small compared to the 2-derivative terms. So again there is no obvious sign of any breakdown of effective field theory. Once again the reason can be traced to general covariance restricting the form of possible higher derivative terms.

\section{Extremal Reissner-Nordstr\"{o}m} \label{ERN}

\subsection{Einstein-Maxwell-scalar theory} \label{EMs theory}

Consider an Einstein-Maxwell-scalar theory where the scalar field is massless and minimally coupled. This theory is described by the action\footnote{We work in units $G=c=1$.}.
\bea
S_{2} = \frac{1}{16 \pi} \int d^4 x \sqrt{-g} \left[ R - F^{\mu \nu} F_{\mu \nu} - \nabla_{\mu} \Phi \nabla^{\mu} \Phi   \right] \, .
\label{einstein maxwell action}
\eea
where $F=dA$ with $A$ a 1-form potential. We now consider higher derivative corrections to this two derivative action. We write the action as
\bea
S = \sum_{k=2}^\infty S_k
\label{full action}
\eea
where $S_2$ is as above and
\bea
S_{k} = \frac{\alpha^{k-2}}{16\pi} \int d^4 x \sqrt{-g} \, {\cal L}_{k}
\label{higher-derivative corrections}
\eea
where $\alpha$ has dimensions of length and ${\cal L}_k$ is a scalar function of the metric, Maxwell field strength and scalar field, involving $k$ derivatives of the scalar field, metric or electromagnetic potential. We will assume that the scalar field is coupled only through its derivatives so the theory possesses a shift symmetry $\Phi \rightarrow \Phi + {\rm const}$. Furthermore, we assume that ${\cal L}_{k}$ does not involve any terms which are linear in (derivatives of) $\Phi$, which implies that setting $\Phi={\rm const}$ is a consistent truncation of the theory.

Since it is not possible to construct a scalar Lagrangian with 3 derivatives, we have $S_3=0$ and the first higher derivative term in the action is $S_4$.

\subsection{Aretakis instability in 2-derivative theory}

\label{aretakis_review}

First we review the Aretakis instability in the 2-derivative theory.
Setting $\Phi={\rm constant}$, the two-derivative theory admits the extreme RN black hole as a solution. We write the metric as
\be
\label{ERNmetric}
 ds^2 = - \delta^2 dv^2 + 2 dv dr + r^2 d\Omega^2 \qquad \delta = 1-\frac{Q}{r}
\ee
and the Maxwell field is
\be
\label{Qdef}
 F = Q d\Omega
\ee
where $d\Omega$ is the volume element on a unit radius $S^2$. We have assumed that the black hole is magnetically charged with charge $Q$.\footnote{We choose magnetically rather than electrically charged BHs for simplicity, as \eqref{Qdef} remains exact under higher derivative corrections. We do not expect any significant differences in the electric case.}

In this background, Aretakis considered linear perturbations in the scalar field, which we write as $\psi \equiv \delta \Phi$. The equation of motion for $\psi$ in the 2-derivative theory is
\be
 \Box \psi = 0.
\ee
We can decompose $\psi$ in spherical harmonics:
\bea
\psi = \sum \psi_{\ell m}(v,r) \, Y_{\ell m}(\Omega) \, ,
\label{spherical harmonic decomposition}
\eea
Because of the spherical symmetry we can ignore the dependence on $m$ and just write $\psi_\ell$. The wave equation becomes
\bea
2r \pa_v \pa_r (r \psi_{\ell}) + \pa_r ((r \d)^2 \pa_r \psi_{\ell}) - \ell (\ell+1) \psi_{\ell} = 0 \, \, .
\label{eom ERN}
\eea
Consider first $\ell=0$. Evaluating  \eqref{eom ERN} at the horizon $\d=0$ shows that the quantity
\bea
H_0 \equiv Q^{-1} \left. \partial_{r} (r \psi_{0}) \right|_{\mathrm{horizon}} \,
\eea
is conserved along the horizon (independent of $v$), and in particular does not decay, for generic initial data, at late times. $H_0$ is called an {\it Aretakis constant}. Since $\left. \psi_0 \right|_{\mathrm{horizon}}$ itself does decay at late times on the horizon \cite{Aretakis:2011ha}, this shows that the first derivative  $\left. \partial_{r} \psi_0 \right|_{\mathrm{horizon}}$ does not decay -- instead, it tends to $H_0$. Higher derivatives of $\psi_0$ behave even `worse' on the horizon: at late times they grow indefinitely, as can be seen by acting on equation \eqref{eom ERN} with $\partial_r$ and restricting to the horizon giving
\bea
Q \pa_v \pa_r^2 \left. (r \psi_0) \right|_\mathrm{horizon} = - H_0 \, .
\label{equation for 2nd derivative}
\eea
Integrating with respect to $v$ then gives
\bea
\pa_r^2 \left. (r \psi_0) \right|_\mathrm{horizon} \sim - \frac{H_0}{Q} v
\label{2 derivative behaviour}
\eea
as $v \to \infty$. It follows that
\be
 \left. \pa_r^2 \psi_0 \right|_\mathrm{horizon}  \sim -\frac{H_0}{Q^2}v
\ee
This can be extended by induction to an arbitrary number of radial derivatives. Acting with $\pa^{k-1}_r$ on \eqref{eom ERN}, restricting to the horizon and integrating along it, shows that
\bea
\pa_r^k \left. \psi_0 \right|_\mathrm{horizon} \sim H_0 Q^{2-2k} v^{k-1}
\label{general number of derivatives}
\eea
as $v \to \infty$, where here and below we ignore dimensionless constants on the RHS. Hence higher derivatives of $\psi_0$ grow polynomially with $v$ at late time on the event horizon. This is the Aretakis instability.

Similar behaviour occurs for $\ell>0$. Acting on \eqref{eom ERN} with $\pa_r^\ell$ and restricting to the horizon shows that there is a conserved quantity
\bea
H_\ell \equiv \frac{1}{Q^2} \pa_r^{\ell} \left[ r \pa_r (r \psi_{\ell}) \right]
\eea
As in the $\ell=0$ case, an inductive procedure yields, for $k \ge \ell+1$
\bea
\pa_r^k \left. \psi_{\ell} \right|_\mathrm{horizon} \sim H_\ell Q^{2(\ell+1-k)} v^{k-1-\ell}
\label{general l and derivatives}
\eea
at late time along the event horizon. Notice that $\ell+2$ derivatives are required to construct a quantity that grows along the horizon, hence the Aretakis instability is strongest for the $\ell=0$ mode.

We will also need to know the behaviour of quantities which decay along the horizon. Numerical results in Ref. \cite{Lucietti:2012xr} strongly suggest that $\psi_0 \sim v^{-1-\ell}$ at least for $\ell=0,1$. This is confirmed by rigorous results of Ref. \cite{aretakis_new}, which prove that \eqref{general l and derivatives} holds for any $k \ge 0$ when the Aretakis constant $H_\ell$ is non-zero. It is also proved that $v$-derivatives behave in the way one would expect by naively differentiating w.r.t. $v$: \be
\pa_v^j \pa_r^k \left. \psi_{\ell} \right|_\mathrm{horizon} \sim v^{k-j-\ell - 1-\epsilon(j,k,\ell)}
\label{gen_deriv}
\ee
where
\be
\epsilon(j,k,\ell) =
\left\{ \begin{array}{ll} 0 & {\rm if} \; k \le \ell \; {\rm or} \; k \ge j + \ell + 1 \\ 1& {\rm if} \; \ell+1 \le k \le j+\ell \end{array} \right.
\label{eps_def}
\ee
We have dropped all coefficients on the RHS of \eqref{gen_deriv}. These coefficients are all proportional to $H_\ell$ multiplied by appropriate powers of $Q$.

Although the following will not be used in our analysis, it is interesting to note that the above late-time behaviour is reproduced by an expression of the form
\bea
r \, \p_\ell = v^{-1-\ell} f^{(\ell)}(v \d) \, ,
\label{ansatz for 0th order wavefunction l>0}
\eea
where $f^{(\ell)}$ is a smooth function with $f^{(\ell)}(0) \ne 0$. This Ansatz can be substituted into (\ref{eom ERN}). Taking the late time $v \to \infty$ limit, keeping $z \equiv v \d$ fixed, (\ref{eom ERN}) then reduces to an ordinary differential equation for $f$. Solving it gives the 0th order wavefunction ($Q=1$):
\bea
r \p_{\ell} = v^{-1-\ell} \left[ \frac{c_{1\ell}}{(2+z)^{\ell+1}} + c_{2\ell} \, \, z^{\ell+1}  {}_2 F_1[1,2\ell+2;\ell+2;-z/2] \right] \, \, ,
\label{0th order solution l>0}
\eea
where $c_i$ are constants. For $\ell=0$, it reduces to
\bea
r \p_0 = \frac{c_{20}}{v} + \frac{H_0}{v(2+v \d)} \, \, ,
\label{0th order solution l=0}
\eea
The late time behaviour here involves {\it two} constants $H_0$ and $c_{20}$. The interpretation of the latter is as a {\it Newman-Penrose constant} \cite{Newman:1968uj}. Just as the Aretakis constants are associated to outgoing radiation propagating along the future event horizon, the NP constants are associated to ingoing radiation propagating along future null infinity. In other words, they correspond to late time {\it ingoing} radiation. In equation (\ref{gen_deriv}) we assumed vanishing NP constants but this result can be generalized to allow non-zero NP constants \cite{aretakis_new}. Henceforth we will assume vanishing NP constants.

\subsection{Higher derivative corrections in near horizon geometry} \label{near horizon geometry ERN}

Setting $\Phi={\rm constant}$, the two-derivative theory admits the extreme RN black hole as a solution. We assume that this solution can be corrected so that it remains a solution of the theory to all orders in $\alpha$. We will assume that the corrected black hole is magnetically charged with charge $Q$ defined by (\ref{Qdef}). Of course this satisfies $dF=0$.

The near horizon geometry of this black hole will be $AdS_2 \times S^2$ where the $AdS_2$ and $S^2$ have radii $L_1$ and $L_2$ respectively. We can write $L_i = Q\tilde{L}_i (\alpha/Q)$ $i=1,2$ where $\tilde{L}_i$ is dimensionless. For small $\alpha/Q$ the higher derivative corrections will be negligible and the $AdS_2$ and $S^2$ will both have radius $Q$. The higher derivative corrections start at ${\cal O}(\alpha^2)$ hence we have
\be
\tilde{L}_1(0) = 1+ {\cal O}(\alpha^2/Q^2) \qquad \tilde{L}_2(0) = 1+ {\cal O}(\alpha^2/Q^2)
\ee
We write the $AdS_2 \times S^2$ metric in ingoing Eddington-Finkelstein coordinates as
\be
\label{ads2corr}
 ds_2^2 = L_1^2 \left( -r^2 dv^2 + 2 dv dr  \right)+ L_2^2 d\Omega^2
\ee
Ref.  \cite{Lucietti:2012xr} showed that a massless scalar in this geometry exhibits the Aretakis instability at the future Poincar\'e horizon $r=0$. At first this seems rather surprising given that a scalar field in $AdS_2 \times S^2$ exhibits no instability in global coordinates. This was discussed in Ref.  \cite{Lucietti:2012xr}, we will expand a little on this discussion here.

For a well-posed problem we need to impose boundary conditions at infinity in $AdS_2$. Following Ref. \cite{Lucietti:2012xr}, we assume that boundary conditions have been chosen such that, in a neighbourhood of $r=0$, $v \rightarrow \infty$ (where the Poincar\'e horizon intersects infinity), these conditions correspond to "normalizable" boundary conditions for the scalar field.

The Aretakis instability does not involve the growth of some scalar quantity, but is instead associated to the growth of the components of a tensor, specifically the second derivative of $\psi$. But how does one know that this growth is associated to some physical effect rather than to bad behaviour of the basis in which the components are calculated? The point is that the asymptotically flat black hole solution has a canonically defined Killing vector field $V$ which generates time translations. One can choose a basis to be time-independent, i.e., Lie transported w.r.t. $V$. If a component of some tensor exhibits growth in such a basis then one can be sure that this is a physical effect rather than an artifact of the choice of basis. An example of such a basis is a coordinate basis where $V$ is one of the basis vectors. This is the case in Eddington-Finkelstein coordinates where $V = \partial/\partial v$. This is why one can be sure that the Aretakis instability is not a coordinate effect.

Now in $AdS_2 \times S^2$ there is a difference because there are different choices that can be made for the generator of time translations. If one chooses a basis invariant under {\it global} time translations then one would not see any instability in higher derivatives of $\psi$. However, we are interested in $AdS_2 \times S^2$ because it arises as the near-horizon geometry of an asymptotically flat black hole. In the near-horizon limit, one obtains not global $AdS_2$ but $AdS_2$ in {\it Poincar\'e} coordinates, and the generator of time translations reduces to $V=\partial/\partial v$, the generator of time translations in the Poincar\'e patch. Hence if one views $AdS_2 \times S^2$ as describing the near-horizon geometry of a black hole then one should use $V$ as the generator of time translations, and choose a basis that is Lie transported w.r.t. $V$. In such a basis the Aretakis instability is present, so the near-horizon geometry captures the behaviour present in the full black hole solution.

Since the Aretakis instability can be seen in the near-horizon geometry, we will start by investigating the effect of higher derivative corrections on this instability in the $AdS_2 \times S^2$ background (\ref{ads2corr}). We will take into account two sources of higher-derivative corrections: first we are using the exact, higher-derivative corrected, background (\ref{ads2corr}). Second, we will include the effect of {\it linear} higher derivative corrections to the scalar field equation of motion. The reason for restricting to linear higher derivative corrections is that if we allow nonlinearity then we have to incorporate the effects of the backreaction of the scalar field on the geometry. However, even in the 2-derivative theory, it is known that this backreaction destroys the $AdS_2$ asymptotics \cite{Maldacena:1998uz}. To incorporate this backreaction we have to consider the full black hole solution, as we will do in the next section.

Since the action does not contain terms linear in $\Phi$, the higher derivative corrections to the Einstein equation and the Maxwell equation also do not contain terms linear in $\Phi$, and the corrections to the scalar equation of motion do not contain any $\Phi$-independent terms. Furthermore, our assumption of a shift symmetry implies that the equations involve only derivatives of $\Phi$. This structure implies that when we linearize around an exact background solution with $\Phi={\rm const}$, the linear perturbation to $\Phi$ decouples from the linear metric and Maxwell field perturbations.

To discuss linear higher-derivative corrections to the scalar field equation of motion we will work at the level of the action. We expand the action to quadratic order in $\psi=\delta \Phi$. We then substitute in the expansion in spherical harmonics \eqref{spherical harmonic decomposition}, and perform the integral over $S^2$. Modes corresponding to different harmonics will decouple from each other, giving an effective action for the field $\psi_{\ell m}$ in $AdS_2$ of the form\footnote{Since the spherical harmonics are complex, it is convenient to allow our scalar field $\psi$ and the fields $\psi_{\ell m}$ to be complex.}
\be
\label{ads2action}
 S_{\ell m} = \int d^2 x \sqrt{-g_2} \sum_{n=0}^\infty c_{\ell  n} \bar{\psi}_{\ell m} \Box^n \psi_{\ell m}
\ee
where $g_2$ is the $AdS_2$ metric (with radius $L_1$), $\Box$ is the d'Alembertian of this metric, and $c_{\ell  n}$ are (real) constants depending on $\alpha$ and $Q$. The form of this effective action is dictated by the $AdS_2$ symmetry of the background. Recall our assumption that the scalar field is derivatively coupled. Derivatives can act on either the $S^2$ or $AdS_2$ directions. But the spherically symmetric $\ell=0$ mode is constant on $S^2$ hence it cannot appear without $AdS_2$ derivatives in the above action. It follows that $c_{00}=0$.

Terms in the action with $n \ge 2$ must arise from higher derivative terms in the original action and hence must appear with appropriate powers of $\alpha$. We can write
\be
 c_{\ell  n} = \alpha^{2n-2} \tilde{c}_{\ell  n}(\alpha/Q) \qquad n \ge 2
\ee
where $\tilde{c}_{\ell  n}$ is a dimensionless function of $\alpha/Q$. For $n=0,1$ we can separate out the terms present in the 2-derivative theory from those arising from the higher derivative corrections (to both the background and the equation of motion):\footnote{We are not bothering to keep track of the overall normalization of the action, i.e., it may differ by a multiplicative constant from that defined in (\ref{higher-derivative corrections}).}
\be
 c_{\ell  0} = -\frac{\ell(\ell+1)}{Q^2} + \frac{\alpha^2}{Q^4} \tilde{c}_{\ell  0} (\alpha/Q)
\ee
\be
 c_{\ell  1} = 1 + \frac{\alpha^2}{Q^2} \tilde{c}_{\ell  1} (\alpha/Q)
\ee
Again $\tilde{c}_{\ell  n}$ is a dimensionless functions of $\alpha/Q$ and $\tilde{c}_{00}=0$.

A standard result in effective field theory is that the lowest order (i.e. two derivative) equation of motion can be used to simplify the higher derivative terms in the action. This is achieved via a field redefinition \cite{burgess}. To see how this works here, perform a field redefinition (here we suppress the $\ell,m$ indices throughout)
\be
 \psi = \hat{\psi} + \sum_{n=2}^\infty \alpha^{2n-2} d_{n}\Box^{n-1} \hat{\psi}
\ee
where the dimensionless coefficients $d_{n}(\alpha/Q)$ are to be determined. We substitute this into the action and let $E_n$ be the coefficient of $\bar{\hat{\psi}} \Box^n \hat{\psi}$. We demand that $E_n=0$ for $n \ge 2$. This gives a set of equations that can be solved order by order in $\alpha/Q$ to determine the coefficients $d_n$. To lowest order, $E_2=0$ fixes $d_2=-\tilde{c}_2/2 + {\cal O}(\alpha^2/Q^2)$. Using this, $E_3=0$ fixes the ${\cal O}(1)$ part of $d_3$. Plugging the latter back into $E_2=0$ then determines the ${\cal O}(\alpha^2/Q^2)$ part of $d_2$. One then uses $E_4=0$ to determine $d_4$ to ${\cal O}(1)$, plug this back into $E_3=0$ to determine $d_3$ to ${\cal O}(\alpha^2/Q^2)$ and then $E_2=0$ determines $d_2$ to ${\cal O}(\alpha^4/Q^4)$. Repeating this process to all orders gives
\be
 S = \int d^2x \sqrt{-g} \left( c_0 \bar{\hat{\psi}} \hat{\psi} + c_1' \bar{\hat{\psi} }\Box \hat{\psi} \right)
\ee
where $c_1' = c_1 + 2(\alpha/Q)^2 c_0 d_2=1+{\cal O}(\alpha^2/Q^2)$. Hence, reinstating $\ell,m$ indices, the equation of motion of $\hat{\psi}_{\ell m}$ is
\be
 \left( \Box - m_\ell^2 \right) \hat{\psi}_{\ell m}=0
\ee
where
\be
m_\ell^2 = -\frac{c_{\ell  0}}{c_{\ell 1}'} = \frac{\ell(\ell+1)}{Q^2} + {\cal O}(\alpha^2/Q^4)
\ee
 so we can write
\be
 m_\ell^2 L_1^2 = \ell(\ell+1) M_\ell(\alpha^2/Q^2)
\label{shifted mass 2}
\ee
for some function $M_\ell$ with $M_\ell(0)=1$. Hence, to all orders in $\alpha$, $\hat{\psi}_{\ell m}$ behaves as a massive scalar field in $AdS_2$ with mass $m_\ell$. Since $\psi_{\ell m}$ is linearly related to $\hat{\psi}_{\ell m}$, the same will be true for $\psi_{\ell m}$. We see that the only effect of the higher derivative corrections is to correct the mass of this scalar field. Of course, all we have done here is to perform a Kaluza-Klein reduction of the scalar field $\psi$ on $S^2$.

Note that the higher derivative corrections do not generate a mass for $\psi_{00}$. The masslessness of $\psi_{00}$ is protected by the assumed shift symmetry, which implies $c_{00}=0$ and hence $m_0^2=0$ to all orders. So {\it higher derivative corrections do not change the equation of motion for the $\ell=0$ mode.}

Now we can discuss the effect of the higher derivative corrections on the Aretekis instability in $AdS_2 \times S^2$. In the absence of such corrections, this instability is strongest in the $\ell=0$ sector, with $\partial_r^2 \psi_{00}$ growing linearly with $v$ along the horizon at $r=0$. For higher partial waves more derivatives are required to see the instability: $\partial_r^{\ell+2} \psi_{\ell m}$ grows linearly with $v$. From the results just obtained, we see that higher derivative corrections have no effect on the $\ell=0$ sector and so $\partial_r^2 \psi_{00}$ will still grow linearly with $v$. However, these corrections do affect higher $\ell$ modes through the change in the mass just discussed. To understand the effect of this change in the mass, we can use results of Ref. \cite{Lucietti:2012xr}, which determined the behaviour of massive scalar fields in $AdS_2$ along the Poincar\'e horizon at late time.\footnote{
To obtain these results it is necessary to assume, as above, that the scalar field obeys "normalizable" boundary conditions in a neighbourhood of where the Poincar\'e horizon intersects infinity.} The result is that, for a scalar of mass $m$, at late time along the horizon $r=0$
\be
\label{kDelta}
 \partial_r^k \psi \propto v^{k-\Delta}
\ee
where $\Delta$ is the conformal dimension
\be
 \Delta = \frac{1}{2} + \sqrt{m^2 L_1^2 + \frac{1}{4}}
\ee
with $L_1$ the $AdS_2$ radius. So for a massive scalar, $\partial_r^k \psi$ decays along the horizon if $k<\Delta$ and grows if $k>\Delta$. Applying this in our case, writing $M_\ell = 1 + \delta M_\ell$ with $\delta M_\ell = {\cal O}(\alpha^2/Q^2)$ we have
\be
 \Delta = \ell+1 + \frac{\ell(\ell+1)}{2\ell+1} \delta M_\ell + \ldots
\ee
If $\delta M_\ell > 0$ then the higher derivative corrections have led to increased stability in the sense that the decay is slightly faster for $k<\ell+1$ and the blow up is slightly slower for $k> \ell+1$. On the other hand, if $\delta M_\ell < 0$ then the higher derivatives lead to reduced stability in the sense that not only do we have faster growth for $k>\ell+1$, we also have power law growth for $k=\ell+1$. In particular, if $\delta M_1<0$ then the second derivative of the $\ell=1$ mode exhibits power law growth along the horizon. However, the exponent in this power law will be proportional to $-\delta M_1$ and therefore small compared to the linear growth exhibited by the second derivative of the $\ell=0$ mode. So even though higher derivative corrections may strengthen the instability in the higher $\ell$ modes, for small $\alpha/Q$, they do not strengthen them enough that they compete with the dominant $\ell=0$ mode, which is unaffected by these corrections.

Of course, the question of whether $\delta M_\ell$ is positive or negative is the same as the question of how higher derivative corrections affect the masses of Kaluza-Klein harmonics when we reduce on $S^2$. In particular, in a theory with sufficient supersymmetry one might expect that $\delta M_\ell \ge 0$ for all modes.

In summary, we have shown that higher derivative corrections to the geometry and linear higher derivative corrections to the scalar field equation of motion do not lead to a qualitative change in the behaviour of linear scalar field perturbations at the Poincar\'e horizon of $AdS_2 \times S^2$. The dominant $\ell=0$ Aretakis instability is protected by the assumed shift symmetry of the scalar field. Higher derivative corrections can lead to small changes in the exponents of the power-law behaviour exhibited by higher $\ell$ modes but, for small $\alpha/Q$, these corrections are small and so the $\ell=0$ instability remains dominant. There is no sign of any breakdown of effective field theory.

Why do the higher derivative corrections to the equation of motion not become large? The reason can be traced to the fact that these corrections appear only via $\Box^n \psi$ in
 (\ref{ads2action}). This structure is a consequence of general covariance, i.e., the fact that the higher derivative terms do not depend on anything except the background geometry. The high degree of symmetry of the background geometry then greatly restricts the form of the higher derivative terms in the action. Note in particular that general covariance forbids the appearance in the action of higher derivative terms evaluated in some geometrically preferred basis, such as the basis (Lie transported w.r.t. $V$) that is used to exhibit the instability.

\subsection{Full black hole solution}

\label{high_deriv_size}

We have just seen that the higher derivative corrections do not cause a problem during the Aretakis instability in the near-horizon geometry. However, as we have just argued, this may be a consequence of the high degree of symmetry of the near-horizon geometry. It is not obvious that this result will still hold if we consider the less symmetric extremal RN geometry. Furthermore, the above analysis did not incorporate nonlinear corrections to the equations of motion (except via correcting the background geometry). In this section we will address both of these deficiencies by considering higher derivative corrections during the Aretakis instability in the full extreme RN geometry.

We will assume that the extremal RN solution can be corrected to give a static, spherically symmetric, solution to all orders in $\alpha$, with $\Phi={\rm const}$. For a large black hole, i.e., one with $\alpha/Q \ll 1$, the effect of corrections to this background solution should be small so we will neglect them in this section. We will focus on the effect of the higher derivative corrections to the equations of motion during the Aretakis instability. For effective theory to remain valid, these terms should remain small, giving perturbative corrections to the 2-derivative theory. If the higher derivative terms become larger than the 2-derivative terms then effective field theory breaks down. So in this section we will investigate whether or not this is the case. We will consider all of the equations of motion, not just the scalar field equation of motion.

First we note that coupled gravitational and electromagnetic perturbations of the extreme RN black hole exhibit an Aretakis instability  \cite{Lucietti:2012sf} but this is weaker than the massless scalar field instability in the sense that it requires more derivatives to see it. So we will continue to focus on the Aretakis instability driven by a massless scalar field. This instability is strongest in the spherically symmetric $\ell=0$ sector. So if higher derivatives are going to cause trouble it seems very likely that this will occur in the $\ell=0$ sector. Therefore we can simplify by restricting to spherical symmetry.

We recall the effect of {\it nonlinearities} in the 2-derivative theory. As discussed in the Introduction, the nonlinear evolution of the spherically symmetric instability in the 2-derivative theory was studied in Ref. \cite{Murata:2013daa}, where it was shown that the initial perturbation can be fine-tuned so that the metric "settles down" to extreme RN on and outside the event horizon, with the scalar field on the horizon exhibiting the Aretakis instability. In other words, the ``most unstable" behaviour exhibited by the nonlinear 2-derivative theory is to give a spacetime which, at late time, looks like a linear scalar field on a fixed extreme RN background.

Motivated by these results, our strategy in this section will be to consider a spherically symmetric scalar field evolving in a fixed extreme RN background. We will perform a consistency check on the smallness of the higher derivative corrections to the equations of motion. To do this we will take the known results for the late time behaviour of the scalar field along the horizon in the 2-derivative theory, and use this to estimate the size of higher derivative corrections to the equation of motion. In particular, we can compare the size of the higher derivative terms to (possibly nonlinear) terms present in the 2-derivative theory. In order for effective field theory to remain valid, the higher derivative terms must remain small compared to the 2-derivative terms.

The extremal Reissner-Nordstrom solution is a type D solution, i.e., the Weyl tensor has two pairs of coincident principal null directions, which are also principal null directions of the Maxwell field. It is convenient to employ the Geroch-Held-Penrose (GHP) formalism \cite{Geroch:1973am}, which is well suited to situations in which one has a pair of preferred null directions. This formalism is based on a null tetrad and enables all calculations to be reduced to the manipulation of scalar quantities.
In the metric (\ref{ERNmetric}) we choose a null tetrad $\{l,n,m,\bar{m}\}$ based on the principal null directions:
\bea
l^a &=& (1,\d^2/2,0,0)  \, , \non \\
n^a &=& (0,- 1,0,0) \, , \non \\
m^a &=& \frac{1}{\sqrt{2} r} \, \left(0,0,1,\frac{i}{\sin\theta}\right) \, ,
\label{tetrad def}
\eea
In the GHP formalism, there is a freedom to change the basis \eqref{tetrad def} so that the two null directions are preserved. One possibility is to rescale the null vectors (referred to as a boost)
\bea
l \to \lambda l \, \, ; \, \, n \to \lambda^{-1} n \, ,
\label{boost definition}
\eea
where $\lambda$ is a real function. The other is to rotate the spatial basis vectors (referred to as a spin)
\bea
m \to e^{i \theta} m  \, \, ; \, \, \bar{m} \to e^{-i \theta} \bar{m} \, .
\label{spin definition}
\eea
where $\theta$ is a real function. Any tensor can be decomposed in the basis \eqref{tetrad def}, the different components then become functions of definite \emph{boost/spin weight}. A function $\eta$ with boost weight $b$ and spin weight $s$, under a combination of \eqref{boost definition} and \eqref{spin definition}, transforms as
\bea
\eta \, \to \, \lambda^{b} e^{i\theta s} \eta \, .
\label{boost+spin transformation}
\eea
The GHP formalism is designed to maintain convariance under boosts and spins. A privileged role is played by objects which transform covariantly, i.e., objects with definite boost and spin weight. Not all connection components transform covariantly. Those that do take the following values in the extreme RN background:
\bea
  &\kappa=\kappa'=\sigma=\sigma'=\tau=\tau'=0& \non \\
 &\rho=-\d^2/(2r) \qquad \rho' = 1/r &
\eea
The GHP scalars $\rho,\rho'$ have boost weights $1,-1$ respectively, and both have zero spin.

Since the background spacetime is type D, the only non-zero components of the Weyl tensor and Maxwell field are those with vanishing boost and spin weights
\bea
\Psi_2 &\equiv & C_{\mu \nu \rho \sigma} l^\mu m^\nu n^\rho \bar{m}^\sigma = -\frac{Q \d}{r^3}\, , \non \\
\phi_1 &\equiv & \frac{1}{2} F_{\mu \nu} \left( l^\mu n^\nu + \bar{m}^\mu m^\nu \right) = -i\frac{Q}{2r^2} \, .
\label{NP nonzero scalars}
\eea
The non-vanishing Ricci tensor components have boost weight zero and are determined by $\phi_1$.

The GHP formalism introduces derivative operators with definite spin/boost weights. In the extreme RN background, they are given by
\bea
\t \, \eta &=& (l^\mu \nabla_\mu - 2 b \,  \epsilon) \, \eta = \left(\pa_v + \frac{\d^2}{2} \, \pa_r - b \, \frac{Q \d}{r^2} \right) \eta  \, \, , \non \\
\t' \eta &=& (n^\mu \nabla_\mu - 2 b \, \gamma) \, \eta  = - \pa_r  \, \eta\, \, ,  \non \\
\e \, \eta &=&  (m^\mu \nabla_\mu - 2 s \, \beta) \, \eta = \frac{1}{\sqrt{2}r} \left( \partial_\theta - s \cot \theta +\frac{i}{\sin \theta} \partial_\phi \right) \eta \, ,  \non \\
\e' \, \eta &=& (\bar{m}^\mu \nabla_\mu + 2 s \, \beta) \, \eta = \frac{1}{\sqrt{2}r} \left( \partial_\theta + s \cot \theta -\frac{i}{\sin \theta} \partial_\phi \right) \eta \, ,
\label{boost weighted operators}
\eea
where $\eta$ is a GHP scalar with boost weight $b$ and spin $s$, and $\epsilon$, $\gamma$ and $\beta$ are Newman-Penrose spin coefficients. The operators $\t$, $\t'$ have zero spin and carry boost weight $1$, $-1$ respectively, and the operators $\e$, $\e'$ have zero boost weight and carry spin $1$, $-1$ respectively.

Finally we will need to use commutators of these derivative operators. Acting on a quantity of boost weight $b$ and spin $s$, in the extreme RN background these are given by
\bea
\left[\t,\t'\right] &=& -2b \left( \Psi_2 + 2|\phi_1|^2 \right) \, , \non \\
\left[\t,\e\right]  &=& \rho \, \e  \, , \non \\
\left[\e,\e'\right] &=& -2s \left( -\rho \rho' - \Psi_2 + 2|\phi_1|^2 \right) \, .
\eea
Now we return to considering the higher-derivative corrected equations of motion in the extreme RN spacetime with a dynamical spherically symmetric scalar field. Consider a boost-weight $B$ component of one of the equations of motion. We will determine the $v$-dependence of higher derivative corrections to this component on the horizon at late time. In the GHP formalism, all quantities are written as scalars so any higher-derivative term can be written in the form $XZ$ where $X$ is constructed entirely from the background GHP scalars and their derivatives, and $Z$ is constructed entirely from the scalar field and its derivatives. We can write $Z=Z_1 \ldots Z_N$ where each $Z_i$ consists of GHP derivatives acting on $\Phi$. Spherical symmetry implies that none of these derivatives can be $\e$ or $\e'$. To see this, note that any $Z_i$ can be written as $\tilde{D}_1 \ldots \tilde{D}_p \e D_1 \ldots D_q \Phi$, or the corresponding expression with $\e$ replaced by $\e'$, where $\tilde{D}_i \in \{\t,\t',\e,\e'\}$ and $D_i \in \{\t,\t'\}$, for some $p,q \ge 0$. But $D_1 \ldots D_q \Phi$ has spin $0$, so, using spherical symmetry, it is annihilated by $\e$ and $\e'$. Hence any $Z_i$ involving $\e$ or $\e'$ must vanish.

Next, using the commutator $[\t,\t']$, we can order $\t$ and $\t'$ derivatives in $Z_i$ so that $\t$ derivatives appears to the left of $\t'$ derivatives. So there is no loss of generality in assuming that each $Z_i$ has the form $\t^j \t'^k \Phi$. Recall that we assumed that $\Phi$ is derivatively coupled but one might wonder whether commutators could generate terms without GHP derivatives. However this is not possible:
$[\t,\t']$ acting on derivatives of $\Phi$ gives a result involving derivatives of $\Phi$ whereas $[\t,\t']$ acting on $\Phi$ gives zero (because $\Phi$ has zero boost weight). Hence commutators cannot give rise to terms involving $\Phi$ without derivatives so we must have $j+k \ge 1$.

Now on the horizon we have $\delta=0$ so we can replace $\t$ with $\partial_v$ in $\t^j \t'^k \Phi$ and converting \eqref{gen_deriv} to GHP notation gives
\be
 \t^j \t'^k \Phi |_\mathrm{horizon}  \sim v^{k-1-j-\epsilon} = v^{-b-1-\epsilon}
\ee
where $b=j-k$ is the boost weight of this term and $\epsilon \in \{0,1\}$ with $\epsilon=0$ if $k=0$ or $k \ge j+1$ and $\epsilon=1$ otherwise. Taking a product of $N$ such terms gives
\be
 Z|_\mathrm{horizon}= \left[ \left( \t^{j_1}  \t^{'k_1} \Phi \right) \ldots \left( \t^{j_N}  \t^{'k_N} \Phi \right) \right] |_\mathrm{horizon} \sim v^{-(b_1+\epsilon_1) - \ldots -(b_N+\epsilon_N) -N}=v^{B_X - B-N -E}
\ee
where $E = \sum \epsilon_i$ and we have used the fact that $XZ$ has boost weight $B$, so we have $\sum b_i =B-B_X$ where $B_X$ is the boost weight of $X$.

Now, since $X$ is constructed from background quantities, it is independent of $v$ hence we have
\be
\label{hi_d_scaling}
XZ|_\mathrm{horizon} \sim v^{B_X-B-N-E}
\ee
We will now show that if $B_X>0$ then $X$ vanishes on the horizon. The scalar $X$ can be written as $X=X_1 \ldots X_M$, where each $X_i$ consists of GHP derivatives acting on some GHP scalar $\omega$ associated to the background spacetime, i.e., $\omega \in \{\rho,\rho',\Psi_2,\phi_1,\phi_1^*\}$. Note that all of these quantities have zero spin and are spherically symmetric. This means that we can argue as above to show that $\e$ or $\e'$ derivatives cannot appear in $X_i$. Using commutators, we can assume that $X_i$ has the form $ \t^j \t'^k  \omega$.  Furthermore, since we can replace $\t$ by $\partial_v$ on the horizon, and the GHP scalars are all $v$-invariant, the expression $\t^j \t'^k \omega$ vanishes when evaluated on the horizon unless $j=0$. So any $X_i$ that is non-vanishing on the horizon must be of the form $\t'^k \omega$. This has boost weight $b_\omega-k$ where $b_\omega$ is the boost weight of $\omega$. Note that the possible $\omega$ all have non-positive boost weight, with the exception of $\omega=\rho$. So if $\omega$ is anything except $\rho$ then $X_i$, if non-vanishing on the horizon, must have non-positive boost weight. If $\omega$ is $\rho$ then $b_\omega=1$ but, since $\rho$ vanishes on the horizon, we need $k \ge 1$ to construct a non-vanishing expression. Hence $X_i$ also has non-positive boost weight in this case. Therefore we have proved that all $X_i$ that are non-vanishing on the horizon must have non-positive boost weight.
This proves that if $X$ is non-vanishing on the horizon then $B_X \le 0$.

Let's apply this to the Einstein equation, which has components with $|B| \le 2$. (Note that spherical symmetry implies that the $B=\pm 1$ components are trivial.) In the 2-derivative theory, the RHS of the Einstein equation involves the energy-momentum tensor of the scalar field. We'll denote this 2-derivative energy momentum tensor as $T_{\mu\nu}^\Phi$. Equation \eqref{gen_deriv} implies that a boost weight $B$ component of $T_{\mu\nu}^\Phi$ scales as $v^{-B-2}$ at late time along the horizon. Hence in order for a higher-derivative term \eqref{hi_d_scaling} to become large compared to the 2-derivative term in a component of boost weight $B$ we would need $B_X-B-N-E>-B-2$, i.e., $B_X>N+E-2$. But we've just seen that non-vanishing $X$ on the horizon requires $B_X \le 0$ so we'd need $N<2-E$ for our higher derivative term to dominate. However, we've assumed that all terms in the action are at least quadratic in the scalar field, which implies that all terms in the Einstein equation have $N \ge 2$ (or $N=0$ but the latter don't depend on the scalar field and hence don't depend on $v$). Hence it is not possible for higher derivatives to become large compared to the 2-derivative terms in the Einstein equation. The "worst" that can happen is that the higher derivative terms exhibit the same scaling with $v$ as the 2-derivative terms. This happens when $N=2$, $E=0$ and $B_X=0$. Such terms scale in the same way as the 2-derivative terms but they will be suppressed by powers of the small quantity $\alpha/Q$.

The same argument can be applied to the scalar field equation of motion, which has $B=0$. A typical 2-derivative term in this equation of motion is $\t \t' \Phi \sim v^{-2}$. So for a higher derivative term to dominate we would need $B_X - N-E > -2$ i.e., $B_X >N+E-2$ so again we'd need $N < 2-E$ for consistency with $B_X \le 0$. Our assumption that the scalar field appears at least quadratically in the action implies that $N \ge 1$ in the scalar field equation of motion. There is now a non-trivial solution to these inequalities given by $N=1$, $E=0$ and $B_X=0$. However, such terms are excluded by our assumption of a shift symmetry. To see this, note that with $N=1$, $Z$ is linear in the scalar field, i.e., of the form $\t^j\t'^k\Phi$ and with $B=B_X=0$ this term must have boost weight $j-k=0$ so $j=k$. Now $E=0$ implies $\epsilon=0$ which is only possible if $j=k=0$, i.e., there are no derivatives acting on $\Phi$. However we explained above that such a term is forbidden by our assumption that the scalar field has a shift symmetry. So in fact the "worst" terms are ones for which the higher derivative terms exhibit the same $v^{-2}$ scaling as the two-derivative terms but are suppressed by powers of $\alpha/Q$. Such terms can have either $N=1$ or $N=2$. With $N=1$ these terms have $Z$ of the form $\t \Phi$ or $\t^j \t'^j \Phi$ with $j \ge 1$. With $N=2$ these terms have $Z$ of the form $(\t^{j_1} \Phi)( \t^{j_2} \t'^{j_1+j_2} \Phi$) with $j_1 \ge 1$, $j_2 \ge 0$.

For the Maxwell equation, it is not possible to compare the $v$-dependence of the higher derivative and 2-derivative terms  because, in spherical symmetry, the Maxwell field does not exhibit any dynamics in the 2-derivative theory, even including nonlinearity. (This is because the scalar field is uncharged.) We can regard the higher derivative corrections as a source term for the Maxwell equation, i.e., as an electromagnetic current. From the above results, a boost weight $B$ component of the current behaves as $v^{B_X -B -N-E}$ at late time on the horizon. Since $B_X \le 0$ and $N \ge 2$ (for the same reason as for the Einstein equation), the most dangerous terms are those with $B_X=0$ and $N=2$, $E=0$, which scale as $v^{-B-2}$. Since components of the Maxwell equation have $|B| \le 1$ we see that these terms decay at late time along the horizon.

These calculations demonstrate that there is no obvious failure of effective field theory on the horizon at late time. Although certain higher derivatives of the scalar field become large on the event horizon at late time, this does not imply that higher derivative corrections to the equation of motion become large compared to the 2-derivative terms. This is because, in the equations of motion, the ``bad" derivatives are always multiplied by "good" terms which are decaying, or by terms $X$ which vanish on the horizon. The reason for this can be traced back to general covariance. This implies that the quantities $X$ appearing in the higher derivative terms are constructed only from GHP scalars associated to the background solution. In particular $X$ depends only on the background fields and not on any additional structure such as a preferred basis. So, just as we found for the near-horizon geometry, it is general covariance which prevents a breakdown of effective field theory.

\section{Extremal Kerr} \label{EK}

In this section we will discuss the scalar field instability at the horizon of an extremal Kerr black hole, first discussed by Aretakis in the axisymmetric case and extended to the non-axisymmetric case in Ref. \cite{Casals:2016mel}. Our goal is to understand whether higher derivative corrections could become important during this instability. As for extremal RN, we will start by analyzing this in the near-horizon geometry before turning to the full black hole solution.

\subsection{Near-horizon analysis}\label{Decay/blow up rates from symmetry} \label{near horizon geometry EK}

As explained above, the near-horizon $AdS_2 \times S^2$ geometry of an extremal RN black hole provides a simplified setting in which to study the Aretakis instability \cite{Lucietti:2012xr}. Here we will consider the near-horizon extremal Kerr (NHEK) geometry \cite{Bardeen:1999px} as a simplified setting to study the Aretakis instability of extremal Kerr. In fact our main motivation here is to go beyond the (axisymmetric) Aretakis instability and consider {\it non-axisymmetric} perturbations of extremal Kerr, as discussed in Ref. \cite{Casals:2016mel}.

In the axisymmetric case, the results of Ref. \cite{Casals:2016mel}  do not see the dominant Aretakis instability, behaving as in (\ref{derivative behaviour on horizon intro}). This is because the approach of Ref. \cite{Casals:2016mel}  cannot incorporate the presence, in the initial data, of outgoing radiation at the event horizon, so all the Aretakis constants are zero. Under such circumstances there is still an instability but it requires an extra derivative to see it \cite{Aretakis:2012bm}, and this "subleading" instability was reproduced in Ref. \cite{Casals:2016mel}. For non-axisymmetric perturbations, Ref. \cite{Casals:2016mel}  found an instability stronger than that discovered by Aretakis, with the first derivative of the scalar field generically growing along the horizon. However, since the approach of Ref. \cite{Casals:2016mel} cannot model outgoing radiation initially present at the event horizon one might wonder whether the inclusion of such radiation would make the non-axisymmetric instability even worse. This is something that we can investigate using the methods of this section.

We will assume that the extremal Kerr solution with $M \gg \alpha$ can be corrected to all orders in $\alpha$ to give an extremal black hole solution of the theory (\ref{full action}) and that this corrected solution has vanishing Maxwell field and constant scalar field. The general results of Ref. \cite{Kunduri:2007vf} imply that the near-horizon geometry of this black hole has $SL(2,R) \times U(1)$ symmetry and the metric can be written as an $S^2$ fibred over $AdS_2$:
\bea
ds^2 = \Lambda_1^2(\a,\theta)\left[-R^2 dT^2 + \frac{dR^2}{R^2}\right] + \Lambda_2^2(\a,\theta) d\theta^2 + \Lambda_3^2(\a,\theta) (d\varphi + k R dT)^2  \, ,
\label{NHEK}
\eea
where $k(\a)$ is a constant and $\Lambda_i$ are smooth functions on the sphere parameterized by $(\theta,\varphi)$. For the uncorrected theory $\a=0$ we recover the NHEK geometry, for which  \cite{Bardeen:1999px}
\bea
k(0)=1 \qquad \Lambda_1^2(0,\theta)=\Lambda_2^2(0,\theta) = M^2 (1+\cos^2\theta) \,, \quad \Lambda_3^2(0,\theta) = 2M^2\sin\theta\, .
\label{gamma and lambda}
\eea
The coordinates $\{T,R,\varphi\}$ are then the near horizon descendants of the time, radial and axial coordinates of extreme Kerr in Boyer-Lindquist form. For nonzero $\a$, we will refer to (\ref{NHEK}) as the $\a$-NHEK geometry.

\begin{figure}[h!]
	\centering
	
	\resizebox{0.3\textwidth}{!}
	{
		
\begin{tikzpicture}
\draw[thick] (0,-5) -- (0, 9);
\draw[thick] (-4,-5) -- (-4, 9);

\draw[thick] (0,-4) to (-4, 0) to (0,4);

\draw[thick,dashed] (0,1.7) to (-3.15, 4.85);

\draw[thick] (0,0) to (-4, 4) to (0,8);

\draw (1,4.3) node[rotate=0]{$T=0$};
\draw (1,3.7) node[rotate=0]{$v\to\infty$};

\draw [thick,dotted] (0,4) circle (7pt);

\draw (-2,-1.5) node[rotate=-45]{$r=0$};
\draw (-2.75, 0.75) node[rotate=45]{$r=0$};

\draw (-2.75,3.25) node[rotate=-45]{$R=0$};
\draw (-2, 5.5) node[rotate=45]{$R=0$};
\draw (-0.25, 5.75) node[rotate=90]{$R=\infty$};

\end{tikzpicture}

	}
	
	\caption{Penrose diagram illustrating the coordinate transformation \eqref{coord transformation}. The coordinates $\{ T,R,\varphi \}$ cover the upper triangular patch. The coordinates $\{ t,r,\chi \}$ cover the lower triangular patch. The point $T=0$, $R=\infty$ or $r=0$, $v \equiv t-1/r \to \infty$, on which we focus, is indicated by the dotted circle. The dashed line is an example for a possible initial data surface.}
	\label{figure}
\end{figure}
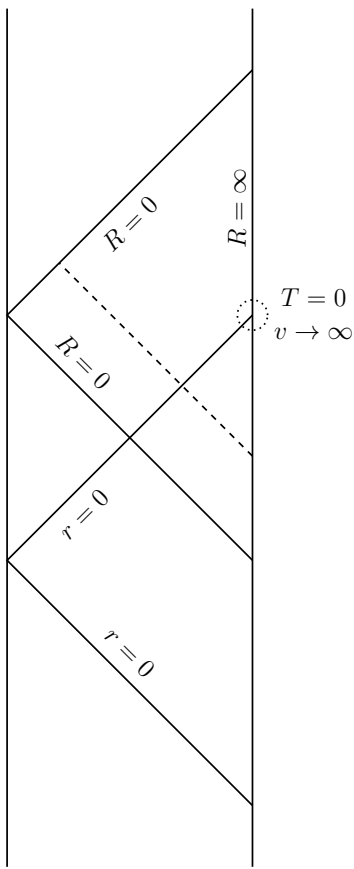

The coordinates $\{T,R,\theta,\varphi\}$ cover a patch of $\a$-NHEK which is analogous to the Poincar{\'e} patch in $\mathrm{AdS_2}$. We can covert to global coordinates (described in appendix \ref{Global NHEK}) to obtain what we will call the global $\a$-NHEK geometry. The $AdS_2$ part of this geometry is depicted by the infinite vertical strip in figure \ref{figure}. One of the $SL(2,R)$ generators of the isometry group can be taken to be the translations in global time $\tau$ (see appendix \ref{Global NHEK}), that is---shifts up and down the `global $\alpha$-NHEK' strip in figure \ref{figure}. We will make use below of a translation with $\D \tau = \pi/2$ which in Poincar\'e corresponds to the transformation (see also \cite{Hadar:2015xpa},\cite{Hadar:2016vmk})
\bea
T &=& -\frac{r^2 t}{r^2 t^2 - 1} \, , \non \\
R &=& \frac{r^2 t^2 - 1}{r} \, ,  \non \\
\varphi &=& \chi + k \log \frac{r t+1}{r t-1} \, .
\label{coord transformation}
\eea
\eqref{coord transformation} is an isometry: the metric in the new coordinates is precisely of the same form as \eqref{NHEK}, replacing $\{T,R,\varphi\} \to \{t,r,\chi\}$.

We will start by considering the wave equation in the above geometry, i.e. we neglect higher derivative corrections to the scalar equation of motion in this section. Supposing initial data for $\p$ is specified on some surface in the near-horizon region, for example $T-1/R = \mathrm{const.}<0$ as seen in figure \ref{figure}, we would like to study the resulting solution.

Ref. \cite{Durkee:2010ea} studied perturbations of near-horizon geometries of the $\a$-NHEK type, and in particular it was shown that they are separable and the wave equation reduces to the equation of a massive charged scalar in $\mathrm{AdS_2}$ with a homogeneous electric field. To see this, use the ansatz
\bea
\p = X(T,R) Y(\varphi,\theta) \, ,
\label{decomposition}
\eea
and Fourier decompose along the $\phi$ direction as
\bea
Y(\varphi,\theta) = e^{i m \varphi} S(\theta) \, .
\label{fourier phi}
\eea
Define the effective $AdS_2$ metric and gauge field
\bea
ds^2 = -R^2 dT^2 + \frac{dR^2}{R^2} \, , \qquad A=-RdT \, .
\label{ads2 metric}
\eea
and the corresponding gauge-covariant derivative
\bea
{\cal D} := \tilde{\nabla} -iqA \, ,
\label{gauge covariant derivative}
\eea
where $\tilde{\nabla}$ is the covariant derivative on $\mathrm{AdS_2}$ and $q=-mk$ is the effective electric charge. Then the equation governing $X(T,R)$ is
\bea
\left( {\cal D}^2 - \lambda - q^2 \right) X(T,R) = 0 \, ,
\label{ads wave equation}
\eea
where $\lambda$ is the eigenvalue of the angular equation
\bea
\mathcal{O}Y := \hat{\nabla}_a \left( \Lambda_1^2 \hat{\nabla}^a Y \right) +q^2 \Lambda_1^2  Y = -\lambda  \, Y\, ,
\label{angular equation aNHEK}
\eea
where $\hat{\nabla}$ is the covariant derivative on the transverse $S^2$ with metric defined by setting $dT=dR=0$ in \eqref{NHEK}. The operator $\mathcal{O}$ can be shown to be self-adjoint w.r.t. an appropriate inner product so its eigenvalues are real and the eigenfunctions form a complete set on $S^2$  \cite{Durkee:2010ea}. Hence there is no loss of generality in decomposing $\p$ as in (\ref{decomposition}). In general, these eigenfunctions can be labelled by a pair of integers $(\ell,m)$ with $|m| \le \ell$ just as for standard spherical harmonics.

Equation \eqref{ads wave equation} describes a scalar field with charge $q$ and squared mass $\mu^2=\lambda+q^2$ in $\mathrm{AdS_2}$ with an electric field. The electric field is homogeneous because the corresponding Maxwell 2-form is proportional to the $AdS_2$ volume form. If one separates variables, i.e., assumes $e^{-i \omega T}$ time dependence then solutions of the radial equation have two possible behaviours as $R \to \infty$, given by \cite{Bardeen:1999px,Amsel:2009ev,Dias:2009ex} $\psi \sim R^{-1/2 \pm (h-1/2)}$ where
\bea
h = \frac{1}{2} + \sqrt{\frac{1}{4} + \lambda} \, .
\label{h def}
\eea
As $R \rightarrow \infty$, a general superposition of such modes will behave as
\bea
X(T,R) =  f_+(T) R^{h-1} \left[1+\mathcal{O}(R^{-1})\right] + f_-(T) R^{-h}\left[1+\mathcal{O}(R^{-1})\right] \, ,
\label{r to infinity approximation}
\eea
for some functions $f_\pm(T)$. For well-defined dynamics we need to impose boundary conditions at $R=\infty$. If $h$ is real then a natural choice is to impose "normalizable" boundary conditions, i.e., $f_+\equiv 0$. In NHEK this is the case for axisymmetric modes, i.e., $m=0$, for which $\lambda = \ell(\ell+1)$ and hence $h = \ell+1$ \cite{Bardeen:1999px}. However, if $\lambda<-1/4$ then $h$ is complex. For NHEK this occurs for non-axisymmetric modes with $|m| \sim \ell$. In this case it is not clear what boundary conditions should be imposed (see Refs.  \cite{Bardeen:1999px,Amsel:2009ev,Dias:2009ex} for discussions of this issue). We will {\it assume} that for complex $h$ one can obtain well-posed dynamics with a boundary condition that fixes some linear relation between $f_+$ and $f_-$.

Notice that the axisymmetric modes will have real $h$ in $\alpha$-NHEK. This is because the associated eigenvalues $\lambda$ are non-negative in NHEK so small higher derivative corrections to the background geometry cannot push $\lambda$ below $-1/4$ in $\alpha$-NHEK. Hence the higher derivative corrections to the background geometry will lead to small real shifts in $h$. This will not happen for the $\ell=0$ mode, i.e., the constant mode on $S^2$, which continues to have $\lambda=0$ and $h=1$ in $\alpha$-NHEK. For the non-axisymmetric modes, it is possible that a NHEK mode with $\lambda$ slightly larger than $-1/4$ (hence real $h$) might correspond to an $\alpha$-NHEK mode with $\lambda$ slightly less than $-1/4$ (hence complex $h$).

The idea now is that we can determine the late time behaviour of the scalar field along the Poincar\'e horizon in $\alpha$-NHEK simply from a coordinate transformation. We consider the Poincar\'e horizon $r=0$ in the coordinates $(t,r,\theta,\chi)$. We shift to ingoing Eddington-Finkelstein coordinates $(v,r,\theta,\chi')$ where
\bea
v=t-\frac{1}{r} \qquad \chi'=\chi-k \log r
\label{ingoing ef}
\eea
so that the metric is now regular at the Poincar\'e horizon:
\bea
ds^2 = \Lambda_1^2(\a,\theta)\left[-r^2 dv^2 + 2dvdr \right] + \Lambda_2^2(\a,\theta) d\theta^2 + \Lambda_3^2(\a,\theta) (d\chi'+ k r dv)^2  \, ,
\eea
Late time along the Poincar\'e horizon corresponds to $r=0$, $v \rightarrow \infty$. From Fig. \ref{figure}, this can be seen to correspond to the limit $R \rightarrow \infty$, $T \rightarrow 0$ in the original coordinates. So we can determine the late-time behaviour of the scalar field by transforming (\ref{r to infinity approximation}) to the new coordinates.
Doing this, including the angular dependence $e^{im\varphi} S(\theta)$, gives
\be
 \psi \approx \left\{ f_+(0) \left[v (rv+2) \right]^{h-1}  + f_-(0) \left[v(rv+2) \right]^{-h} \right\} e^{im\chi'} \left(\frac{rv+2}{v} \right)^{imk} S(\theta)
\ee
Here we have transformed to the new coordinates and taken the limit $v \rightarrow \infty$ with $rv$ fixed. In figure \ref{figure}, $rv$ represents the angle of approach to the center of the dotted circle as the limit $v \to \infty$ is taken. On the horizon we have $rv=0$ but it is convenient to allow for non-zero $rv$ because it enables us to see explicitly the $r$-dependence of $\psi$ at late time near the horizon.

For the modes with real $h$, which includes the axisymmetric modes, we impose normalizable boundary conditions $f_+(0)=0$. From the above expression we have
\be
\label{realh1}
 |\psi|_{\rm horizon} \sim v^{-h}
\ee
and
\be
\label{realh2}
 |\partial_v^j \partial_r^k D^l \psi|_{\rm horizon} \sim v^{k-j-h}
\ee
where $D$ denotes angular derivatives.\footnote{If $h$ is an integer, as for axisymmetric modes in the NHEK geometry, one has to include $\epsilon$ in the exponent as in (\ref{gen_deriv}), (\ref{eps_def}) (replacing $\ell+1$ by $h$). But in $\alpha$-NHEK we do not expect $h$ to be exactly integer except for the $\ell=m=0$ mode, which has $h=1$.} Note that when $h$ is real we have $h \ge 1/2$.

For modes with complex $h$, which are non-axisymmetric, we have $h=1/2+i \zeta$ where $\zeta$ is real. We then have
\be
\label{complexh1}
 |\psi|_{\rm horizon} \sim v^{-1/2}
\ee
and
\be
\label{complexh2}
 |\partial_v^j \partial_r^k D^l \psi|_{\rm horizon} \sim v^{k-j-1/2}
\ee
This is precisely the late time behaviour discovered for the full extremal Kerr solution in Ref. \cite{Casals:2016mel}. As mentioned above, the approach of Ref. \cite{Casals:2016mel} cannot incorporate the effects of outgoing radiation initially present at the event horizon (or non-vanishing Aretakis constants in the axisymmetric case) so one might wonder whether the presence of such radiation could change the results, perhaps leading to even slower decay. Our analysis allows for outgoing radiation initially present at the event horizon and our results agree with those of Ref. \cite{Casals:2016mel} when $h$ is complex. This suggests that inclusion of the initial outgoing radiation does not lead to slower decay. Of course it would be desirable to confirm this using an analysis in the full black hole spacetime rather than just the near-horizon geometry.

The analysis of this section could also be generalised to fields of higher spin, where one would need to supplement the transformation \eqref{coord transformation} with a tetrad rotation (c.f. \cite{Hadar:2014dpa}).

\subsection{Linear higher-derivative corrections in near-horizon geometry}

So far we have studied a massless scalar in the $\alpha$-NHEK geometry, i.e., we have incorporated higher derivative corrections to the background geometry but not to the scalar equation of motion. In this section we will investigate the effects of the {\it linear} higher derivative corrections to the massless scalar equation of motion. We cannot consider nonlinear corrections to the equations of motion because it is known that 2-derivative nonlinearities (i.e. backreaction) tend to destroy the NHEK asymptotics \cite{Amsel:2009ev,Dias:2009ex}.

We will proceed as we did for $AdS_2 \times S^2$ in section  \ref{near horizon geometry ERN}, i.e, expanding the action to quadratic order in $\psi$, substituting in the expansion of $\psi$ in terms of spheroidal harmonics on $S^2$:
\be
 \psi = \sum_{\lambda,m} X_{\lambda m} Y_{\lambda m}
\ee
and then integrating over $S^2$ to obtain an action governing the charged fields $X_{\lambda m}$ in $AdS_2$ with a homogeneous electric field as in (\ref{ads2 metric}). The axisymmetry of the background implies that modes corresponding to harmonics with different values of $m$ will decouple from each other in the action. However, the $\theta$-dependence of the background will lead to coupling of the modes with different values of $\lambda$ (but the same $m$) in the dimensional reduction of the higher derivative terms. Because of the $SL(2,R)$ symmetry of the background, the resulting action for the fields of charge $q=-km$ will have the form (integrating by parts so derivatives act on $X$ and not $\bar{X}$)
\be
\label{NHEKaction}
 S_m = \int d^2 x \sqrt{-g_2} \sum_{\lambda,\lambda',n} c_{m \lambda \lambda' n} \bar{X}_{\lambda m} ({\cal D}^2)^n X_{\lambda' m}
\ee
where $g_2$ is the $AdS_2$ metric in (\ref{ads2 metric}) and (since the action is real)
\be
c_{m\lambda \lambda' n} = \bar{c}_{m\lambda' \lambda n}
\ee
Our assumption that $\psi$ is derivatively coupled implies that $X_{00}$ cannot appear without derivatives in the above action. This is because $Y_{00}$ is constant and hence eliminated by angular derivatives, so $X_{00}$ must be acted on by $AdS_2$ derivatives. Therefore we must have $c_{0\lambda 00}=0$ and hence $c_{00\lambda'0}=0$.

It is convenient to define a vector ${\bf X}_m$ with components $X_{\lambda m}$ and Hermitian matrices ${\bf C}_{mn}$ with components $c_{m\lambda \lambda' n}$. The action can then be written
\be
\label{Sm}
S_m = \int d^2 x \sqrt{-g_2} \sum_n {\bf X}_m^\dagger {\bf C}_{mn} ({\cal D}^2)^n {\bf X}_m
\ee
Since ${\bf C}_{mn}$ is the coefficient of a term with $2n$ derivatives we must have\footnote{Note that the background $AdS_2$ metric in (\ref{ads2 metric}) has unit radius so our coordinates are dimensionless, hence the extra powers of $M$ compared to section \ref{near horizon geometry ERN}. }
\be
 {\bf C}_{mn} = \left( \frac{\alpha}{M} \right)^{2n-2} \tilde{\bf C}_{mn}(\alpha/M) \qquad n \ge 2
\ee
for some dimensionless Hermitian $\tilde{\bf C}_{mn}$. For $n=1,0$ we can use the known equation of motion in the 2-derivative theory and the fact that the higher derivative corrections start at ${\cal O}(\alpha^2)$ to deduce
\be
\label{C1eq}
 {\bf C}_{m1} ={\bf I} + \frac{\alpha^2}{M^2} \tilde{\bf C}_{m1}(\alpha/M)
\ee
and that
\be
{\bf C}_{m0} = {\bf J}_m + \frac{\alpha^2}{M^2} \tilde{\bf C}_{m0}(\alpha/M)
\ee
where ${\bf J}_m$ has components
\be
 j_{m\lambda \lambda'} =  -\left[ \lambda + (mk)^2 \right] \delta_{\lambda \lambda'}
\ee
In the above we are ignoring a possible overall factor in the action.

We now repeat the strategy of section \ref{near horizon geometry ERN} using a field redefinition to eliminate the higher derivative terms in $S_m$. Henceforth we suppress the $m$ index and write
\be
 {\bf X}= \hat{\bf X} + \sum_{n=2}^\infty \left( \frac{\alpha}{M} \right)^{2n-2} {\bf D}_n ({\cal D}^2)^{n-1} \hat{\bf X}
\ee
where ${\bf D}_n$ are dimensionless matrices depending on $\alpha/M$. Substituting this into the action gives
\be
 S = \int d^2 x \sqrt{-g_2} \sum_n \hat{\bf X}^\dagger {\bf E}_n ({\cal D}^2)^n \hat{\bf X}
\ee
where $\hat{\bf X}$ is a vector with components $\hat{X}_\lambda$ and ${\bf E}_n$ are Hermitian matrices. The first few of these are
\be
 {\bf E}_0 = {\bf C}_0 \qquad \qquad {\bf E}_1 = {\bf C}_1 +  \frac{\alpha^2}{M^2} \left( {\bf C}_0 {\bf D}_2 + {\bf D}_2^\dagger {\bf C}_0 \right)
\ee
\be
 {\bf E}_2 = \frac{\alpha^2}{M^2} \left( {\bf C}_1 {\bf D}_2 + {\bf D}_2^\dagger {\bf C}_1 + \tilde{\bf C}_2 \right) + \frac{\alpha^4}{M^4} \left( {\bf C}_0 {\bf D}_3 + {\bf D}_3^\dagger {\bf C}_0 + {\bf D}_2^\dagger {\bf C_0} {\bf D}_2 \right) \, .
\ee
We now want to choose the unknown matrices ${\bf D}_n$ so that ${\bf E}_n$ vanishes for $n \ge 2$. This can be done order by order in $\alpha/M$. We start with ${\bf E}_2 =0$ which, using (\ref{C1eq}), gives ${\bf D}_2 = -\tilde{\bf C}_2/2+{\cal O}(\alpha^2/M^2)$. Then ${\bf E}_3=0$ gives ${\bf D}_3 = -(1/2)\tilde{\bf C}_3 + (3/8)\tilde{\bf C}^2_2 + {\cal O}(\alpha^2/M^2)$. Plugging this back into ${\bf E}_2=0$ then determines the ${\cal O}(\alpha^2/M^2)$ part of ${\bf D}_2$. Repeating this process order by order we achieve ${\bf E}_n=0$ for all $n \ge 2$. The action has become
\be
\label{Snew}
 S = \int d^2 x \sqrt{-g_2} \left( \hat{\bf X}^\dagger {\bf C}_0 \hat{\bf X} + \hat{\bf X}^\dagger {\bf E}_1 {\cal D}^2 \hat{\bf X} \right) \, .
\ee
${\bf E}_1$ is Hermitian so we can diagonalize it with a unitary matrix ${\bf U}$:
\be
 {\bf E}_1 = {\bf U} {\bf K} {\bf U}^\dagger
\ee
where ${\bf K}$ is real and diagonal. Furthermore we have ${\bf E}_1 = {\bf I} + {\cal O}(\alpha^2/M^2)$ so ${\bf K} = {\bf I} + {\cal O}(\alpha^2/M^2)$ and we can choose ${\bf U} = {\bf I} + {\cal O}(\alpha^2/M^2)$. Since ${\bf K}$ is positive definite we can write ${\bf K} = {\bf L}^\dagger {\bf L}$ for a positive definite real diagonal matrix ${\bf L} = {\bf I} + {\cal O}(\alpha^2/M^2)$. We now bring the kinetic term to canonical form with a final field redefinition:
\be
\hat{\bf X}' = {\bf L} {\bf U}^\dagger \hat {\bf X}
\ee
so
\be
 S = \int d^2 x \sqrt{-g_2} \left( -\hat{\bf X}'^\dagger {\bf M}\hat{\bf X}' + \hat{\bf X}'^\dagger  {\cal D}^2 \hat{\bf X}' \right)
\ee
where we have defined the Hermitian "mass matrix"
\be
{\bf M} = -({\bf L}^{-1})^\dagger {\bf U}^\dagger {\bf C}_0 {\bf U} {\bf L}^{-1} =- {\bf J} + {\cal O}(\alpha^2/M^2)
\ee
${\bf M}$ can be diagonalized by a unitary transformation
\be
 {\bf M} = {\bf U}' {\bf M}' {\bf U}'^\dagger
\ee
where ${\bf M}'=-{\bf J} + {\cal O}(\alpha^2/M^2)$ is real and diagonal, and ${\bf U}' = {\bf I} + {\cal O}(\alpha^2/M^2)$. Defining $\hat {\bf X}'' = {\bf U}'^\dagger \hat{\bf X}'$ we finally have decoupled equations of motion:
\be
 {\cal D}^2 \hat{X}''_{\lambda m} - [ \lambda + (km)^2 + {\cal O}(\alpha^2/M^2)]\hat{X}''_{\lambda m}=0
\ee
where we have reinstated the $m$ indices.

We have now included the effects of higher derivative terms both via the correction to the background geometry, and via the correction to the linearized equation of motion for the scalar field. Both effects can be incorporated simply by a perturbative shift $\lambda \rightarrow \lambda + \mathcal{O}(\a^2/M^2)$ in the value of $\lambda$ that appears in the effective $AdS_2$ equation of motion. This translates into a perturbative shift of the conformal weights \eqref{h def} which determine the decay rates at late time along the Poincar\'e horizon.

Recall that the slowest decaying modes are non-axisymmetric with complex $h$, i.e., $\lambda<-1/4$. For these modes, a small perturbative shift in $\lambda$ will still result in complex $h$ and hence the decay results (\ref{complexh1}) and (\ref{complexh2}) will still hold. So we conclude that {\it higher derivative corrections to the background and linear higher derivative corrections to the scalar equation of motion do not change the rate of decay of the slowest decaying NHEK modes}.

For modes with real $h$, the shift in $\lambda$ will result in a small correction to the decay rates (\ref{realh1}), (\ref{realh2}), similar to what happens to the $\ell>0$ modes in $AdS_2\times S^2$, as described in section \ref{near horizon geometry ERN}. However (after field redefinitions) the $\lambda=0$, $m=0$ mode does not suffer a correction, as a consequence of the shift symmetry of the scalar field. To see this, note that $\hat{X}_{00}$ does not appear in the "mass" term in (\ref{Snew}) because of $c_{00\lambda'0} = c_{0\lambda00}=0$. Hence varying (\ref{Snew}) w.r.t. $\hat{X}_{00}$ gives an equation of motion $({\bf E}_1)_{0 \lambda} {\cal D}^2 \hat{X}_{0\lambda}=0$. So $({\bf E}_1)_{0\lambda} \hat{X}_{0\lambda} = \hat{X}_{00} + {\cal O}(\alpha^2/M^2)$ satisfies a decoupled equation of motion with $\lambda=0$.

In summary, our near-horizon analysis, taking into account all higher derivative corrections to the background, and linear higher derivative corrections to the equation of motion, indicates that higher derivative corrections do not make the scalar field instability of Ref. \cite{Casals:2016mel} any worse. So the near-horizon analysis does not indicate any breakdown of effective field theory at late time at the horizon. As for $AdS_2\times S^2$, the reason for this is that general covariance combined with the $SL(2,R)$ symmetry greatly restricts the possible form of the higher derivative terms in the action (\ref{NHEKaction}).

\subsection{Higher derivative corrections in full black hole geometry}

We have shown that higher derivative corrections do not cause a problem during the scalar field instability in the NHEK geometry. However, this may be a consequence of the high symmetry of this near-horizon geometry. It is not obvious that this result will still hold if we consider the less symmetric extremal Kerr geometry. Furthermore, the above analysis did not incorporate nonlinear corrections to the equations of motion (except via correcting the background geometry). In this section we will address both of these deficiencies by considering higher derivative corrections during the scalar field instability in the full extremal Kerr geometry.

We will perform calculations analogous to the calculations we performed for extremal Reissner-Nordstrom in section \ref{high_deriv_size}. We will assume that the extreme Kerr solution can be corrected, to all orders in $\a$, to give a stationary, axisymmetric, neutral BH solution. Assuming that the BH is large, $\a \ll M$, will allow us to neglect the corrections to the background in this section's analysis. We will then take the known behaviour of a massless scalar field on the horizon of an extremal Kerr black hole and use it to compare the size of higher derivative corrections to the equation of motion to the size of two-derivative terms.\footnote{
Note that {\it linearized gravitational} perturbations of extremal Kerr exhibit an Aretakis instability \cite{Lucietti:2012sf}. But this is weaker than the massless scalar instability in the sense that it requires more derivatives to see it. So we will assume that the instability is driven by a massless scalar.}

There is an immediate problem with this investigation. In the two derivative Einstein-scalar theory, there has been no study of backreaction of the scalar field instability of extremal Kerr. So if the effects of two derivative nonlinearities are not understood, how are we to understand higher derivative terms? In this section we will simply {\it assume}, in analogy with the extremal RN case, that the "worst" behavior in the nonlinear two-derivative theory is that the spacetime settles down to extremal Kerr on and outside the event horizon, with the scalar field behaving just like a linear field in the extreme Kerr spacetime. With this assumption, we will determine the behaviour of higher derivative terms in the equations of motion.

We start with the Kerr metric written in ingoing Kerr coordinates $(v,r,\theta,\tilde{\chi})$:
\bea
&ds^2 = -(1-\frac{2Mr}{|\xi|^2}) dv^2 + 2 dv dr -2 M \sin^2 \theta dr d\tilde{\chi} - \frac{4 M^2 r \sin^2 \theta}{|\xi|^2} dv d\tilde{\chi} +\frac{\Sigma}{|\xi|^2} \sin^2 \theta d \tilde{\chi}^2 + |\xi|^2 d \theta^2& \,  . \non \\ \non \\
&\xi = r+iM\cos\theta \qquad  \d=1-M/r \qquad  \Sigma = (r^2+M^2)^2 - M^2 r^2 \d^2 \sin^2 \theta &
\label{line element kerr}
\eea
The event horizon is at $r=M$ i.e. $\delta=0$. We now convert to co-rotating coordinates $(v,r,\theta,\chi)$ defined by
\bea
\tilde{\chi} = \chi + v/2M \, .
\label{corotating coordinates}
\eea
In these coordinates, $\partial/\partial v$ is tangent to the horizon generators. The Kerr solution is type D and we choose a null tetrad based on the two repeated principal null directions. In coordinates $(v,r,\theta,\chi)$, the basis is
\bea
l^a &=& \left( 2(r^2 + M^2),r^2 \d^2,0,\d(M+r^2/M) \right) \,  , \non \\
n^a &=& - \frac{1}{2 |\xi|^2} \, (0,1,0,0) \, \, , \non \\
m^a &=& \frac{1}{\sqrt{2} \xi} \, \left(i M \sin \theta,0,1,\frac{i(1+\cos^2\theta)}{2\sin\theta}\right) \,  ,
\label{tetrad def kerr}
\eea
The GHP connection scalars are:
\bea
&\kappa = \kappa' = \sigma = \sigma' = 0& \non \\
&\tau = \frac{i M \sin \theta}{\sqrt{2} |\xi|^2} \qquad \tau' = \frac{i M \sin \theta}{\sqrt{2} \bar{\xi}^2} \qquad \rho = \frac{r^2 \d^2}{\bar{\xi}} \qquad \rho' = - \frac{1}{2 \bar{\xi}^2 \xi}& \, .
\eea
The type D property means that the only non-vanishing GHP curvature scalar is
\be
 \Psi_2 = -\frac{M}{\xi^3} \, .
\ee
The GHP derivative operators are given by
\bea
\t \, \eta &=& \left[ 2(r^2+M^2)\pa_v+r^2 \d^2 \pa_r+\d(M+r^2/M)\pa_\chi +2b r \d \right] \eta  \, \, ,  \\
\t' \eta &=& \left[  - \frac{1}{2|\xi|^2} \pa_r +\frac{1}{|\xi|^4}\left(  b r+ i s M \cos \theta \right)  \right] \, \eta\, \, ,  \non \\
\e \, \eta &=&  \left[ \frac{1}{\sqrt{2} \xi} \left( i M \sin\theta \pa_v + \pa_\theta+\frac{i(1+\cos^2\theta)}{2\sin\theta} \pa_\chi \right) +s\frac{\cot\theta}{2\xi}-(b-s)\frac{iM\sin\theta}{\sqrt{2}\xi^2} \right] \eta \, ,  \non \\
\e' \, \eta &=&  \left[ \frac{1}{\sqrt{2} \bar{\xi}} \left( -i M \sin\theta \pa_v + \pa_\theta-\frac{i(1+\cos^2\theta)}{2\sin\theta} \pa_\chi \right)-s\frac{\cot\theta}{\sqrt{2}\bar{\xi}}+(b+s)\frac{iM\sin\theta}{\sqrt{2}\bar{\xi}^2} \right] \eta \, , \non
\eea
Commutators of these derivatives acting on a quantity of boost weight $b$ and spin $s$ are given by
\bea
\left[ \t , \t' \right] &=& (\bar{\tau}-\tau')\e+\left(\tau-\bar{\tau'}\right)\e'-(b+s)\left(-\tau \tau'+\Psi_2\right)-(b-s)\left(-\bar{\tau} \, \bar{\tau'}+\bar{\Psi}_2\right) \, , \non \\
\left[ \t , \e \right] &=& \bar{\rho} \, \e - \bar{\tau} \, \t -(b-s)\bar{\rho} \, \bar{\tau'} \, , \non \\
\left[ \e , \e' \right] &=& \left( \bar{\rho'}-\rho' \right)\t+\left( \rho-\bar{\rho} \right)\t' + (b+s)\left(\rho \rho' +\Psi_2 \right) - (b-s)\left(\bar{\rho} \, \bar{\rho'} +\bar{\Psi}_2 \right) \, .
\eea
Consider a component of the equations of motion which has boost weight $B$. As in section \ref{high_deriv_size} we note that any higher derivative term has the form $XZ$ where $X$ is constructed from background GHP quantities and $Z$ is constructed from the scalar field and its derivatives. We write $Z= Z_1 \ldots Z_N$ where each $Z_i$ consists of GHP derivatives acting on $\Phi$. Using GHP commutators we can arrange these derivative so that $Z_i$ has the form $\t^j \t'^k \e^l \e'^m \Phi$. The assumed shift symmetry implies that, before using commutators, $\Phi$ always appears with derivatives acting on it. From the explicit form of the commutators, we see that a commutator acting on derivatives of $\Phi$ gives terms involving derivatives of $\Phi$ and a commutator acting on $\Phi$ also gives derivatives of $\Phi$ (because $\Phi$ has $b=s=0$). Hence commutators cannot generate terms involving $\Phi$ without derivatives so $j+k+l+m \ge 1$.

We assume that $\Phi$ is composed of all possible harmonics in extreme Kerr, so the late time behaviour is dominated by the non-axisymmetric modes with $m \sim \ell$, i.e. the modes with complex $h$, for which, on the horizon at late time \cite{Casals:2016mel}
\be
|\partial_v^j \partial_r^k D^l \Phi |_{\rm horizon} \sim v^{k-j-1/2}
\label{worst modes kerr}
\ee
where $D$ denotes angular derivatives. Since $\t \sim \partial_v$ on the horizon, this implies that $\t^j \t'^k \e^l \e'^m \Phi\sim v^{k-j-1/2} = v^{-b-1/2}$ where $b=j-k$ is the boost weight of this term. From this we have
\be
 Z|_{\rm horizon} \sim v^{-b_1 - \ldots -b_N -N/2} =v^{B_X - B - N/2}
\ee
where $B_X$ is the boost weight of $X$. Since $X$ is constructed from background quantities, it is independent of $v$ so we also have
\be
\label{kerr_hi_d}
 XZ|_{\rm horizon} \sim v^{B_X - B - N/2}
\ee
We will now show that if $B_X>0$ then $X$ vanishes on the horizon. We write $X=X_1 \ldots X_M$ where each $X_i$ consists of GHP derivatives acting on some GHP scalar $\omega$ associated to the background spacetime, i.e., $\omega \in \{\tau,\tau',\rho,\rho',\Psi_2\}$ (or complex conjugates of these). Using commutators we can assume that $X_i$ has the form $\t^j \t'^k \e^l \e'^m \omega$. Since $\t \sim \partial_v$ on the horizon, and all GHP scalars are $v$-invariant, it follows that this expression vanishes on the horizon unless $j=0$. So any $X_i$ that is non-vanishing on the horizon must have the form $\t'^k \e^l \e'^m \omega$, which has boost weight $b_\omega-k$ where $b_\omega$ is the boost weight of $\omega$. Note that $b_\omega \le 0$ unless $\omega = \rho$. So if $\omega \ne \rho$ then $X_i$, if non-vanishing on the horizon, must have non-positive boost weight. If $\omega=\rho$ then $b_\omega=1$ but, since $\rho$ vanishes on the horizon, we need $k \ge 1$ to construct a non-vanishing expression. So in this case, $X_i$ also has non-positive boost weight if non-vanishing on the horizon. It follows that $B_X \le 0$ if $X$ is non-vanishing on the horizon.

Now let's apply this to the Einstein equation, which has components with $|B| \le 2$. In the 2-derivative theory, the energy momentum tensor of $\Phi$ has components which scale as $v^{-B-1}$ at late time along the horizon. So in order for the higher derivative term \eqref{kerr_hi_d} to become large compared to this 2-derivative term we would need $B_X-B-N/2>-B-1$, i.e., $2B_X>N-2$. But non-vanishing $X$ require $B_X \le 0$ so this is possible only if $N < 2$, which contradicts our assumption that the scalar field appears at least quadratically in the action and hence quadratically in the Einstein equation. So it is not possible for the higher-derivative terms to become large compared to the 2-derivative terms. The worst that can happen is for the higher derivative terms to exhibit the same $v$-dependence as the 2-derivative terms, suppressed by powers of the small quantity $\alpha/M$. This happens if $N=2$ and $B_X=0$.

For the scalar field equation of motion we have $B=0$ and typical 2-derivative terms are $\t\t'\Phi \sim \e\e'\Phi \sim v^{-1/2}$. So for a higher derivative term to become large compared to this we would need $B_X- N/2>-1/2$, i.e., $2B_X>N-1$. But $B_X \le 0$ and in the scalar field equation of motion we have $N \ge 1$ so this is not possible. The worst that can happen is when $N=1$ and $B_X=0$, i.e., linear, boost weight zero, higher derivative corrections with $Z$ of the form $\t^j \t'^j \e^l \e'^m \Phi$. These exhibit the same late time $v$-dependence as the 2-derivative terms but they are suppressed by powers of $\alpha/M$.

In summary, our conclusions are the same as for extremal RN. Even though the non-axisymmetric scalar field instability of extremal Kerr is worse than the axisymmetric Aretakis instability, we have found that, at the horizon, higher derivative corrections remain small compared to 2-derivative terms. Once again the underlying reason for this can be traced  to general covariance, which greatly restricts the form of the higher derivative terms. Specifically, it implies that the quantity $X$ in the above argument is constructed from GHP scalars associated to the background geometry. This gave us the restriction $B_X \le 0$ which eliminates dangerous higher derivative terms in the above argument.

We should emphasize that the analysis of this section started from the assumption that, when we include backreaction in the 2-derivative theory, the ``worst" than can happen is that the spacetime ``settles down" to extremal Kerr, with the scalar field evolving at late time as a test field in the extremal Kerr background. If this assumption is incorrect then our analysis would no longer apply. So clearly the most important issue here is to understand this backreaction in the 2-derivative theory.

\section*{Acknowledgements}
We are grateful to Stefanos Aretakis, Mahdi Godazgar, Amos Ori, and Peter Zimmerman for useful discussions and correspondence. SH is supported by the Blavatnik Postdoctoral Fellowship. SH is grateful to the Albert Einstein Institute, Potsdam for hospitality during the completion of this work. Part of this work was completed while HSR was a participant in the "Geometry and Relativity" programme at the Erwin Schr\"odinger Institute, Vienna.

\appendix

\section{Global $\alpha$-NHEK} \label{Global NHEK}
As in the $\mathrm{AdS_2}$ case, \eqref{NHEK} admits an analytic extension via transformation to `global $\alpha$-NHEK' coordinates:
\bea
R &=& \sqrt{1+y^2} \cos \tau + y \, , \nonumber \\ \nonumber \\
T &=& \frac{\sqrt{1+y^2} \sin \tau}{R} \, , \nonumber \\ \nonumber \\
\varphi &=& \tc + k \log\left| \frac{\cos\tau + y\sin\tau}{1 + \sqrt{1+y^2} \sin\tau} \right| \, .
\label{global NHEK coordinates}
\eea
In these coordinates, the metric becomes
\bea
ds^2 = \Lambda_1^2 \left[ -(1+y^2) d\tau^2 + \frac{dy^2}{1+y^2}\right] + \Lambda_2^2 \,  d\theta^2 + \Lambda_3^2  (d\varphi + k y d\tau)^2  \, .
\label{global NHEK}
\eea


\begin{thebibliography}{99}


\bibitem{Dafermos:2016uzj}
  M.~Dafermos, G.~Holzegel and I.~Rodnianski,
  ``The linear stability of the Schwarzschild solution to gravitational perturbations,''
  arXiv:1601.06467 [gr-qc].

\bibitem{Dafermos:2014cua}
  M.~Dafermos, I.~Rodnianski and Y.~Shlapentokh-Rothman,
  ``Decay for solutions of the wave equation on Kerr exterior spacetimes III: The full subextremal case $|a| < M$,''
  arXiv:1402.7034 [gr-qc].

\bibitem{Aretakis:2011ha}
  S.~Aretakis,
  ``Stability and Instability of Extreme Reissner-Nordstr\'om Black Hole Spacetimes for Linear Scalar Perturbations I,''
  Commun.\ Math.\ Phys.\  {\bf 307}, 17 (2011)
  doi:10.1007/s00220-011-1254-5
  [arXiv:1110.2007 [gr-qc]].

\bibitem{Aretakis:2011hc}
  S.~Aretakis,
  ``Stability and Instability of Extreme Reissner-Nordstrom Black Hole Spacetimes for Linear Scalar Perturbations II,''
  Annales Henri Poincare {\bf 12}, 1491 (2011)
  doi:10.1007/s00023-011-0110-7
  [arXiv:1110.2009 [gr-qc]].

\bibitem{Aretakis:2012ei}
  S.~Aretakis,
  ``Horizon Instability of Extremal Black Holes,''
  Adv.\ Theor.\ Math.\ Phys.\  {\bf 19}, 507 (2015)
  doi:10.4310/ATMP.2015.v19.n3.a1
  [arXiv:1206.6598 [gr-qc]].

\bibitem{Aretakis:2012bm}
  S.~Aretakis,
  ``A note on instabilities of extremal black holes under scalar perturbations from afar,''
  Class.\ Quant.\ Grav.\  {\bf 30}, 095010 (2013)
  doi:10.1088/0264-9381/30/9/095010
  [arXiv:1212.1103 [gr-qc]].

\bibitem{Lucietti:2012sf}
  J.~Lucietti and H.~S.~Reall,
  ``Gravitational instability of an extreme Kerr black hole,''
  Phys.\ Rev.\ D {\bf 86}, 104030 (2012)
  doi:10.1103/PhysRevD.86.104030
  [arXiv:1208.1437 [gr-qc]].

\bibitem{Lucietti:2012xr}
  J.~Lucietti, K.~Murata, H.~S.~Reall and N.~Tanahashi,
  ``On the horizon instability of an extreme Reissner-Nordstr\'om black hole,''
  JHEP {\bf 1303}, 035 (2013)
  doi:10.1007/JHEP03(2013)035
  [arXiv:1212.2557 [gr-qc]].

\bibitem{Casals:2016mel}
  M.~Casals, S.~E.~Gralla and P.~Zimmerman,
  ``Horizon Instability of Extremal Kerr Black Holes: Nonaxisymmetric Modes and Enhanced Growth Rate,''
  Phys.\ Rev.\ D {\bf 94}, no. 6, 064003 (2016)
  doi:10.1103/PhysRevD.94.064003
  [arXiv:1606.08505 [gr-qc]].

\bibitem{Gralla:2016sxp}
  S.~E.~Gralla, A.~Zimmerman and P.~Zimmerman,
  ``Transient Instability of Rapidly Rotating Black Holes,''
  Phys.\ Rev.\ D {\bf 94}, no. 8, 084017 (2016)
  doi:10.1103/PhysRevD.94.084017
  [arXiv:1608.04739 [gr-qc]].

\bibitem{Zimmerman:2016qtn}
  P.~Zimmerman,
  ``Horizon instability of extremal Reissner-Nordstr\"om black holes to charged perturbations,''
  Phys.\ Rev.\ D {\bf 95}, no. 12, 124032 (2017)
  doi:10.1103/PhysRevD.95.124032
  [arXiv:1612.03172 [gr-qc]].

\bibitem{Aretakis:2013dpa}
  S.~Aretakis,
  ``Nonlinear instability of scalar fields on extremal black holes,''
  Phys.\ Rev.\ D {\bf 87}, 084052 (2013)
  doi:10.1103/PhysRevD.87.084052
  [arXiv:1304.4616 [gr-qc]].

\bibitem{Angelopoulos:2016hln}
  Y.~Angelopoulos, S.~Aretakis and D.~Gajic,
  ``Asymptotic blow-up for a class of semilinear wave equations on extremal Reissner-Nordstr\"om spacetimes,''
  arXiv:1612.01562 [math.AP].

\bibitem{Murata:2013daa}
  K.~Murata, H.~S.~Reall and N.~Tanahashi,
  ``What happens at the horizon(s) of an extreme black hole?,''
  Class.\ Quant.\ Grav.\  {\bf 30}, 235007 (2013)
  doi:10.1088/0264-9381/30/23/235007
  [arXiv:1307.6800 [gr-qc]].

\bibitem{Dodelson:2015toa}
  M.~Dodelson and E.~Silverstein,
  ``String-theoretic breakdown of effective field theory near black hole horizons,''
  Phys.\ Rev.\ D {\bf 96}, no. 6, 066010 (2017)
  doi:10.1103/PhysRevD.96.066010
  [arXiv:1504.05536 [hep-th]].

\bibitem{aretakis_new}
Y.~Angelopoulos, S.~Aretakis and D.~Gajic, "Late-time asymptotics for the wave equation on extremal Reissner-Nordstrom", to appear.

\bibitem{Newman:1968uj}
  E.~T.~Newman and R.~Penrose,
  ``New conservation laws for zero rest-mass fields in asymptotically flat space-time,''
  Proc.\ Roy.\ Soc.\ Lond.\ A {\bf 305}, 175 (1968).
  doi:10.1098/rspa.1968.0112

\bibitem{Maldacena:1998uz}
  J.~M.~Maldacena, J.~Michelson and A.~Strominger,
  ``Anti-de Sitter fragmentation,''
  JHEP {\bf 9902}, 011 (1999)
  doi:10.1088/1126-6708/1999/02/011
  [hep-th/9812073].

\bibitem{burgess}
C.~P.~Burgess,
  ``Introduction to Effective Field Theory,''
  Ann.\ Rev.\ Nucl.\ Part.\ Sci.\  {\bf 57}, 329 (2007)
  doi:10.1146/annurev.nucl.56.080805.140508
  [hep-th/0701053].

\bibitem{Geroch:1973am}
  R.~P.~Geroch, A.~Held and R.~Penrose,
  ``A space-time calculus based on pairs of null directions,''
  J.\ Math.\ Phys.\  {\bf 14}, 874 (1973).
  doi:10.1063/1.1666410

\bibitem{Bardeen:1999px}
  J.~M.~Bardeen and G.~T.~Horowitz,
  ``The Extreme Kerr throat geometry: A Vacuum analog of AdS(2) x S**2,''
  Phys.\ Rev.\ D {\bf 60}, 104030 (1999)
  doi:10.1103/PhysRevD.60.104030
  [hep-th/9905099].

\bibitem{Kunduri:2007vf}
  H.~K.~Kunduri, J.~Lucietti and H.~S.~Reall,
  ``Near-horizon symmetries of extremal black holes,''
  Class.\ Quant.\ Grav.\  {\bf 24}, 4169 (2007)
  doi:10.1088/0264-9381/24/16/012
  [arXiv:0705.4214 [hep-th]].

\bibitem{Durkee:2010ea}
  M.~Durkee and H.~S.~Reall,
  ``Perturbations of near-horizon geometries and instabilities of Myers-Perry black holes,''
  Phys.\ Rev.\ D {\bf 83}, 104044 (2011)
  doi:10.1103/PhysRevD.83.104044
  [arXiv:1012.4805 [hep-th]].

\bibitem{Hadar:2015xpa}
  S.~Hadar, A.~P.~Porfyriadis and A.~Strominger,
  ``Fast plunges into Kerr black holes,''
  JHEP {\bf 1507}, 078 (2015)
  doi:10.1007/JHEP07(2015)078
  [arXiv:1504.07650 [hep-th]].

\bibitem{Hadar:2016vmk}
  S.~Hadar and A.~P.~Porfyriadis,
  ``Whirling orbits around twirling black holes from conformal symmetry,''
  JHEP {\bf 1703}, 014 (2017)
  doi:10.1007/JHEP03(2017)014
  [arXiv:1611.09834 [hep-th]].

\bibitem{Amsel:2009ev}
  A.~J.~Amsel, G.~T.~Horowitz, D.~Marolf and M.~M.~Roberts,
  ``No Dynamics in the Extremal Kerr Throat,''
  JHEP {\bf 0909}, 044 (2009)
  doi:10.1088/1126-6708/2009/09/044
  [arXiv:0906.2376 [hep-th]].

\bibitem{Dias:2009ex}
  O.~J.~C.~Dias, H.~S.~Reall and J.~E.~Santos,
  ``Kerr-CFT and gravitational perturbations,''
  JHEP {\bf 0908}, 101 (2009)
  doi:10.1088/1126-6708/2009/08/101
  [arXiv:0906.2380 [hep-th]].

\bibitem{Hadar:2014dpa}
  S.~Hadar, A.~P.~Porfyriadis and A.~Strominger,
  ``Gravity Waves from Extreme-Mass-Ratio Plunges into Kerr Black Holes,''
  Phys.\ Rev.\ D {\bf 90}, no. 6, 064045 (2014)
  doi:10.1103/PhysRevD.90.064045
  [arXiv:1403.2797 [hep-th]].

\end{thebibliography}
\end{document}